\documentclass[%
 reprint,
superscriptaddress,
groupedaddress,
unsortedaddress,
runinaddress,
frontmatterverbose, 
 amsmath,amssymb,
 aps,
pra,
floatfix,
10pt]{revtex4-2}

\usepackage{physics}
\usepackage{multirow}
\usepackage{listings}%
\usepackage[colorlinks,linkcolor=blue, citecolor = blue, urlcolor = blue]{hyperref}
\usepackage{physics,wrapfig}
\usepackage{amsmath,graphics}
\usepackage{ragged2e}
\usepackage{adjustbox}
\usepackage{subcaption}
\usepackage{float}
\usepackage{orcidlink}
\usepackage{diagbox}

\begin{document}

\title{Squeezing in conditional measurement setup with coherent input}

\author{E. Devibala \orcidlink{0000-0001-8827-5858}}
\email{devibalaesakkimuthu@gmail.com}

\author{A. Basherrudin Mahmud Ahmed \orcidlink{0000-0001-9194-4653} }\thanks{Corresponding author}
\email{abmahmed.physics@mkuniversity.ac.in}

\affiliation{Department of Theoretical Physics, School of Physics, Madurai Kamaraj University, Madurai, 625021, Tamil Nadu, India }



\begin{abstract}
Conditional Measurement scheme which employs linear optical elements and photon detection is the fertile ground for nonclassical state generation. We consider a simple setup that requires a coherent state and a number state as inputs of the beam splitter, and a photon detector. We show that by tuning the parameters involved in the setup, we can achieve optimal squeezing from the setup. This is facilitated by writing the output state of the conditional measurement as displaced qudits. Setting aside displacement which plays no role in squeezing, the finite-dimensional representation makes it possible to calculate the maximal amount of squeezing. By fixing the detection at one photon level irrespective of any number state input and carefully chosen coherent parameter and beam splitter reflectivity values, one can reach the maximal squeezing at least for lower number state inputs. This is in contrast to the earlier attempts in atom field interaction models etc., where the squeezing obtained was far from saturation. To accommodate the experimental imperfections, we consider the impure nature of the photon source and detector inefficiency.
\end{abstract}


\maketitle

\section{Introduction}
A beam splitter (BS) is a common, no-loss, and inactive optical device in which the two incoming light fields interfere and produce two outgoing fields. In some instances, the BS is used to split the signal field into two parts, and a measurement is performed on one output port with the help of an optical apparatus such as a photodetector and homodyne or heterodyne detector to extract the information encoded in the signal mode. Once the information carried by the signal mode is obtained, the field at the other output port of BS remains unobserved. In 1994, Ban studied the nonclassical properties of this unobserved output state \cite{Ban1994}. The main assumption considered was that the unobserved state must obey the projection postulate of quantum mechanics. The features of this unobserved state are changed because of the effect of the measurement, as the two output modes are correlated to each other. The quantum state of the unobserved mode that is affected by the measurement is referred to as the conditional output state of the BS \cite{Ban1996}. In other words, measurement performed conditionally in one mode of BS output, to generate a quantum state, is called conditional measurement (CM) \cite{Dakna1997}. This method is more beneficial since it includes low-noise photon detectors that are easily accessible. Dakna et al. have done extensive research on CM for the generation of several quantum states, namely Schrodinger-cat-like states \cite{Dakna1998c}, photon-added states \cite{Dakna1998a}, and quantum state engineering \cite{Dakna1998b, Dakna1999a, Dakna1999b}. 

Following them, Lvovsky et al. introduced the quantum optical catalysis (QOC) method, a subclass of CM that generates the superposition of a vacuum state and a single-photon state \cite{Lvovsky2002}. Several protocols using QOC have been developed such as hole burning \cite{Baseia2004}, fock filtration \cite{Sanaka2005} and multiphoton catalysis \cite{Bartley2012}. Hu et al. have investigated multiphoton catalysis as a "Laguerre polynomial excited coherent state" and examined its nonclassical properties \cite{Zubairy2016}. Using a single photon catalysis process, Kruse et al. proposed an experimental protocol for the orthogonalization of a pair of coherent states \cite{Kruse2017}. Recently, displaced qudits (DQ) representations of multiphoton catalysis were studied when a coherent state and photon number states are fed as inputs \cite{Esakkimuthu2024}. The DQ form of the QOC serves as finite superposition of number states preceded by the displacement operator. Moreover, it simplifies the task of identifying the associated states of maximal non-Gaussianity and squeezing that are retained within the output of QOC. In this study, we followed the same methodology in order to express the CM output state within the DQ framework. Similarly, DQ of CM enables the production of arbitrary superpositions of number states with displacement by modifying superposition coefficients.

Squeezing features are needed in nonclassical states to improve the phase precision of detecting weak signals \cite{Loudon1987, Walls1983, Caves2013, Schnabel2017}. Finite-number state superpositions (FSNS) are studied to have a squeezing nature \cite{Wodkiewicz1987, Orllowski1991, Figurny1993}. Initially, the Jaynes-Cummings model was used to generate FSNS \cite{Lee1993, Wodkiewicz1987}. However, the states produced by these models often fell short of achieving optimal squeezing. FSNS can be produced through the quantum scissors technique, which is based on linear optics \cite{Pegg1998}, using two sequential beam splitters, three inputs, and two detectors \cite{Barnett1999}. Within this framework, we mainly focused on inducing squeezing at the CM output when both inputs are in non-squeezed states. We study the amount of squeezing generated in DQ which is generated from CM. One of the major findings is that it is possible to achieve optimal squeezing in the respective number-state superposition created.   

The paper is arranged as follows: In Sect. \ref{sec:DQ}, we discuss the interpretation of the CM output as displaced qudits and elaborated the roles of the superposition coefficients in displaced qubits and displaced qutrits. In Sect. \ref{sec:QS}, we examine the optimal squeezing achieved in DQ with reference to squeezing in the Fock state superpositions. In Sect. \ref{sec:NG}, we analyze the non-Gaussianity of DQ employing the Hilbert-Schmidt distance and the Wigner function. In Sect. \ref{sec:NI} we discuss the experimental implementation of DQ by the fidelity between ideal and nonideal DQ along with its success probability. In Sect. \ref{sec:CN} we summarize our results and drawn the conclusions. 

\section{Displaced Qudits} \label{sec:DQ}
\begin{figure}
\begin{center}
\includegraphics[width=7.5cm,height=3.2cm]{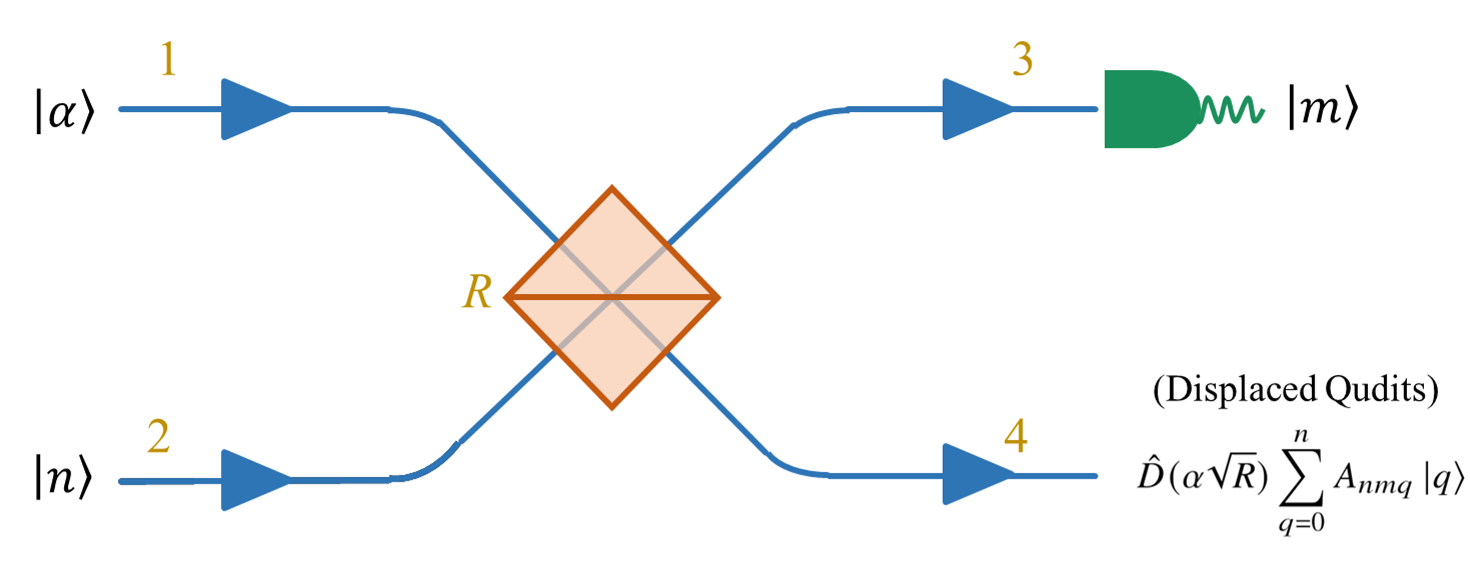}
\caption{The figure shows the schematic representation of CM to generate Displaced Qudits. The pure classical coherent state is given at input port 1 and $\ket{n}$ is sent at input port 2. The output signal $\ket{\psi}_{nm}$ is resulted at output port 4 after the $m$-photon detection at the ancillary port 3.}
\label{Fig.1}
\end{center}
\end{figure}
To implement CM, the input ports 1 and 2 of BS are fed with a coherent state ($\ket{\alpha}$) and an arbitrary photon number state ($\ket{n}$), respectively. After the action of BS, an entangled two-mode state is obtained at the output ports since one of the inputs is nonclassical i.e. $\ket{n}$ is nonclassical. Upon successful projection of some photon number state ($\ket{m}$) on output port 3, the uncorrelated subsystem (a new quantum state) arrives at another output port 4 as shown in Fig. \ref{Fig.1}. The whole process of CM can be derived using,
\begin{equation}
\ket{\psi}_{nm}=\bra{m}\hat{U}_{BS}\ket{n}\ket{\alpha}.
\end{equation}
The unitary operator $\hat{U}_{BS}=\exp \{ \theta(\hat{a}^{\dagger}\hat{b}-\hat{a}\hat{b}^{\dagger}) \}$ describes the action of BS mathematically. Here, $\hat{a}^{\dagger}$ and $\hat{a}$ are photon creation and annihilation operators respectively. After some extent of calculations, the output state of CM can be expressed as \cite{Zhan2022}, 
\begin{align}
    \ket{\psi}_{nm} \sim \frac{\big(-\sqrt{R} \big)^n}{\sqrt{n!m!}} \hbox{H}_{n,m} \Bigg(\alpha \sqrt{1-R}, \hat{a}^{\dagger} \sqrt{\frac{1-R}{R}} \Bigg) \ket{\alpha \sqrt{R}}. 
    \label{Eq:Multi}
\end{align}
Here, $\alpha, R, n$ and $m$ are the input coherent strength, the reflectivity of the BS, the input photon number state and the detected photon number state respectively. We have taken $\theta=\cos^{-1}(\sqrt{R})$ of BS for the condition satisfying $R+T=1$. $\hbox{H}_{n,m} (\bullet)$ represents the two indices of two variable Hermite polynomials and it is represented as \cite{Gorska2019}, 
\begin{align*}
    \hbox{H}_{n,m}(x,y) \overset{\underset{\mathrm{def}}{}}{=} \sum^{min\{n,m\}}_{k=0} \binom{n}{k} \binom{m}{k} (-1)^{k} k! x^{n-k} y^{m-k}.
\end{align*}
Followed by the BS action and conditional measurement, the input states are influenced by two transformations. In the first case, the reduction of the coherent amplitude in the resulting state is achieved. We can see this changes explicitly in Eq. \ref{Eq:Multi}, where the input amplitude reduces to $|\alpha\sqrt{R} \rangle$. Hence, the average photon number may not be conserved, since measurement is involved. In the second place, the $\hbox{H}_{n,m} (\bullet)$ term always possesses $(-1)^k$ which introduces a negative in the odd power terms and induces phase rotation in the system. 

Our purpose is to create a DQ form of CM output. Thus, we shifted the displacement operator in Eq. \ref{Eq:Multi} to the left side of the creation operator $(\hat{a}^{\dagger})$ by considering $|\alpha\sqrt{R} \rangle = \hat{D}(\alpha\sqrt{R}) \ket{0} $. The obtained displaced qudits are expressed as

\begin{align}
    \ket{\psi}_{nm} = \hat{D}(\alpha\sqrt{R}) \sum_{q=0}^{n} A_{nmq} \ket{q}.
    \label{Eq.CMGen}
\end{align}
$\hat{D}(\alpha\sqrt{R})=e^{\alpha\sqrt{R}\hat{a}^\dagger-\alpha^*\sqrt{R}\hat{a}}$ is the displacement operator and $A_{nmq}$ is the superposition coefficients. The above expression reveals the generation of finite superpositions of number states with displacement. In accordance with the input photon number state, $n = 1,2$ and $3$ the output states are called displaced qubits $(\hbox{DQ}_{2})$, displaced qutrits $(\hbox{DQ}_{3})$, and displaced ququarts $(\hbox{DQ}_{4})$.

The main observation is that the input photon states inevitably set a truncation limit to the number of superposition terms created in the output state. In other words, for given $\ket{n}$, the state is truncated to produce up to $n$ superposition terms. For example, if we send a two-photon state to the input port, then the resulted state is $\ket{\psi}_{2m} \sim  \sum_{q=0}^{2} A_{2mq} \ket{q}$ for any detection of photon states. Thus, the detection of photon state simply changes the coefficient amplitudes but does not affect the number of superposition states generated. The superposition coefficients $A_{nmq}$ are computed as,
\begin{align}
    A_{nmq} = N_{nm} \, \binom{n}{q} \, \sqrt{q!} \, \bigg( \frac{1-R}{R}\bigg)^{\frac{q}{2}} \, \hbox{H}_{n-q,m} \big(\alpha\sqrt{1-R}\big).
\end{align}
$N_{nm}$ is the normalization constant. It is calculated as,
\begin{align*}
    N_{nm} = & \Bigg\{ \frac{R^n}{m!n!} \, e^{-\abs{\alpha}^2 (1-R)}  \sum_{q=0}^{n} \, \binom{n}{q}^2 \, q!\, \bigg( \frac{1-R}{R}\bigg)^q \\ & \times \Big[ \hbox{H}_{n-q,m} \big(\alpha\sqrt{1-R}\big)\Big]^2 \Bigg\}^{-\frac{1}{2}}.
\end{align*}
\begin{figure*}[t]
     \begin{subfigure}{0.245\textwidth}
         \includegraphics[scale=0.22]{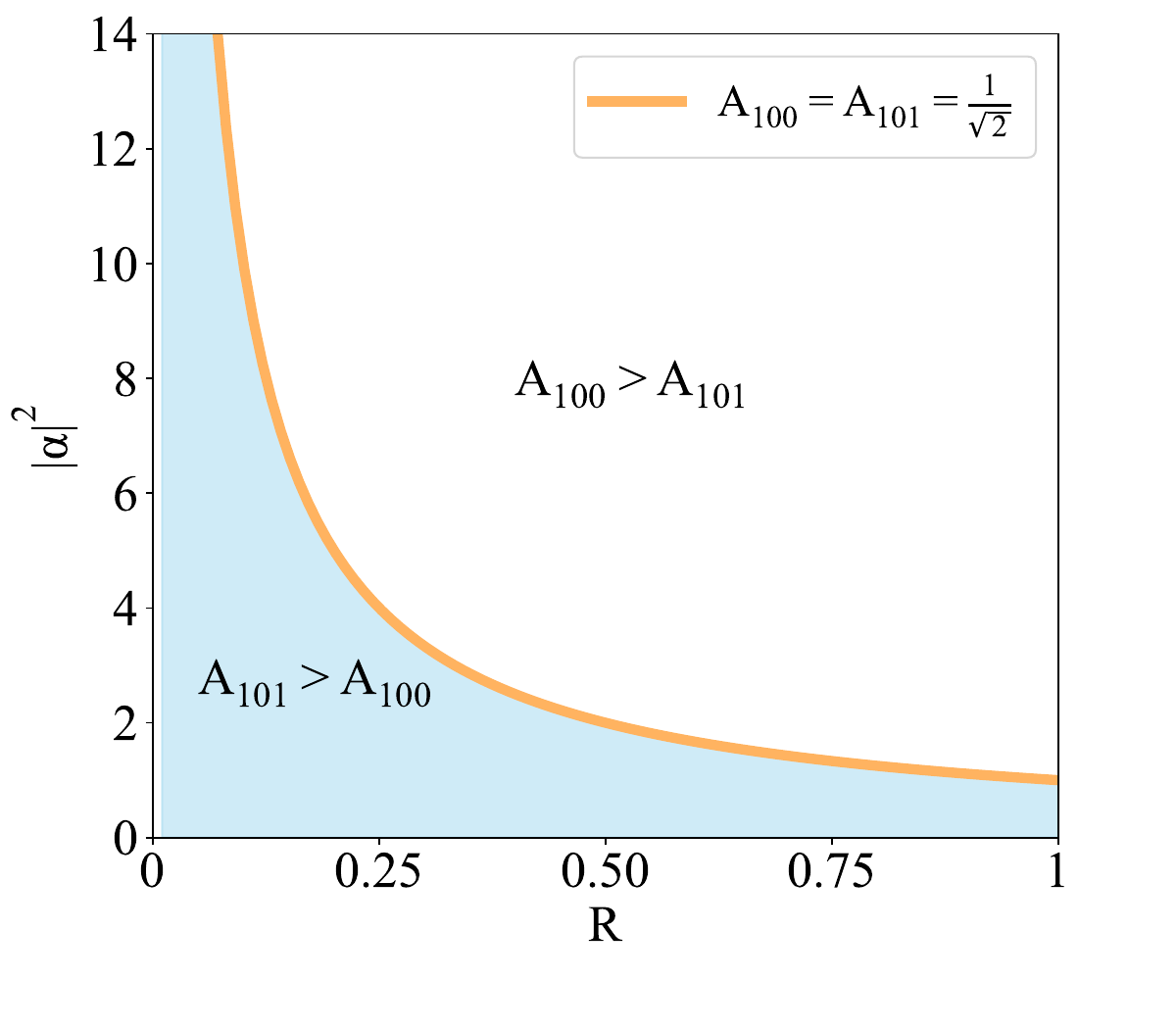}
         \subcaption[]{}
         \label{Fig.A10line}
     \end{subfigure}
     \hfill
     \begin{subfigure}{0.245\textwidth}
         \includegraphics[scale=0.22]{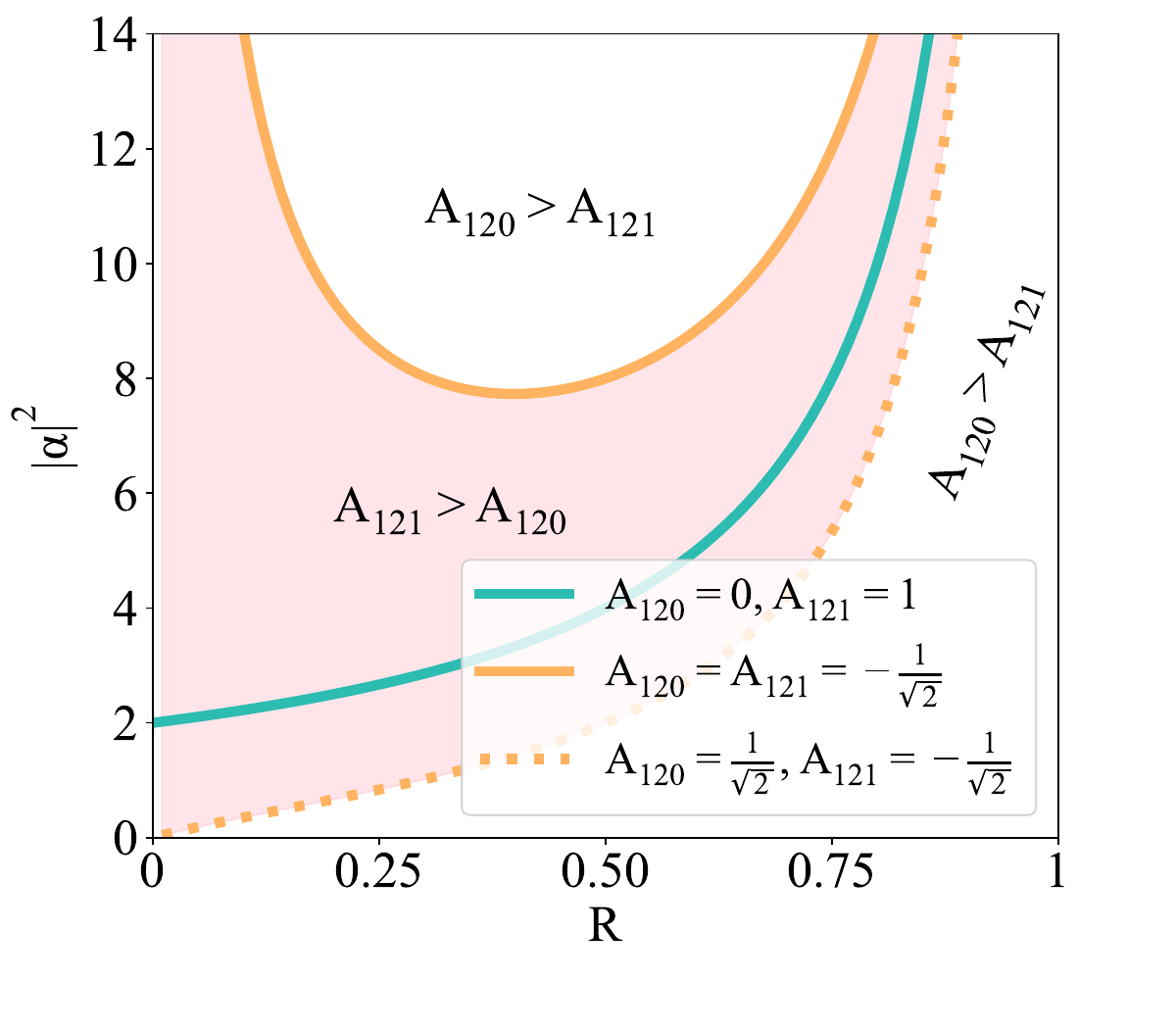}
         \subcaption[]{}
         \label{Fig.A12line}
     \end{subfigure}
     \hfill
     \begin{subfigure}{0.245\textwidth}
         \includegraphics[scale=0.22]{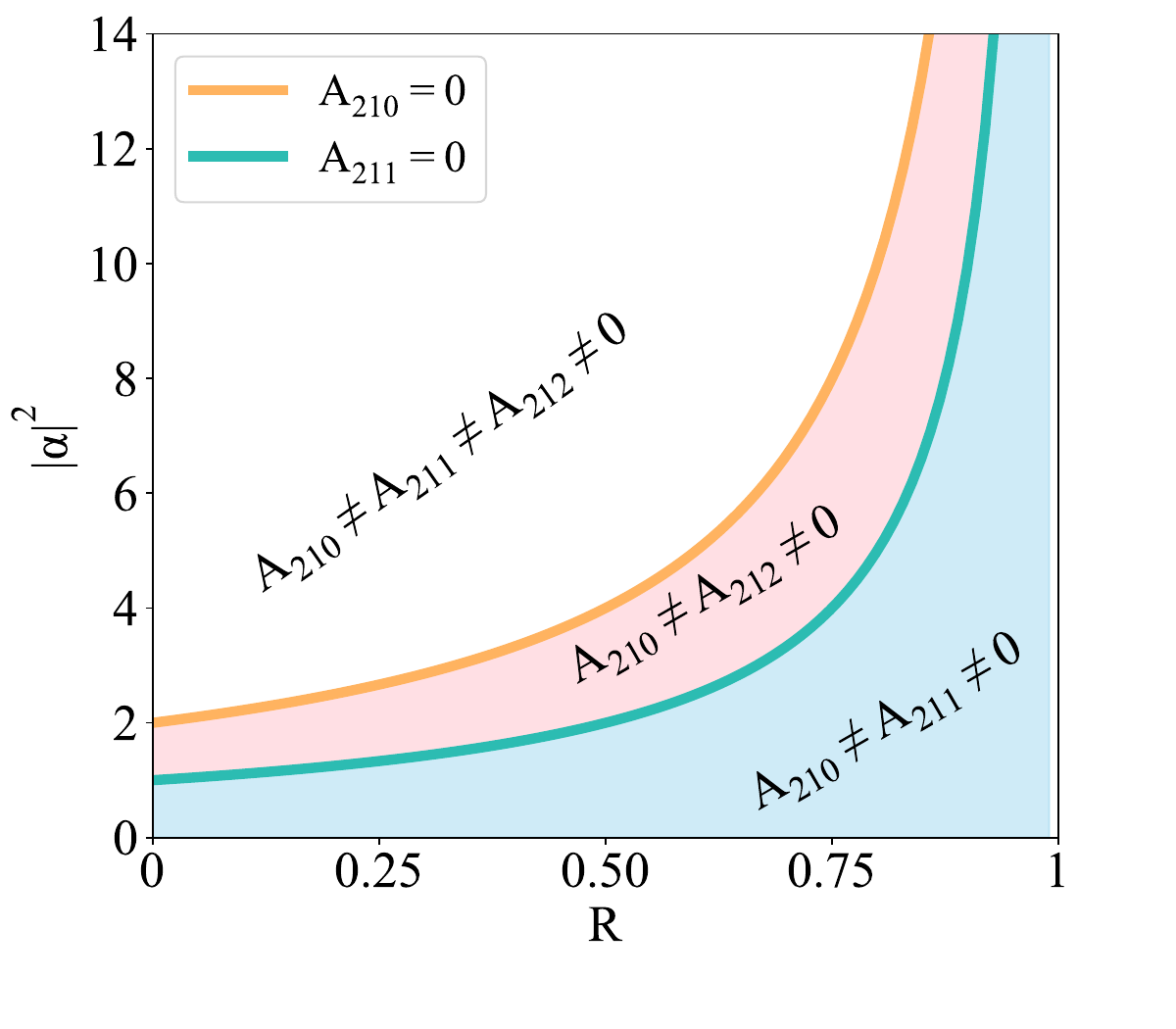}
         \subcaption[]{}
         \label{Fig.A21line}
     \end{subfigure}
      \begin{subfigure}{0.245\textwidth}
         \includegraphics[scale=0.22]{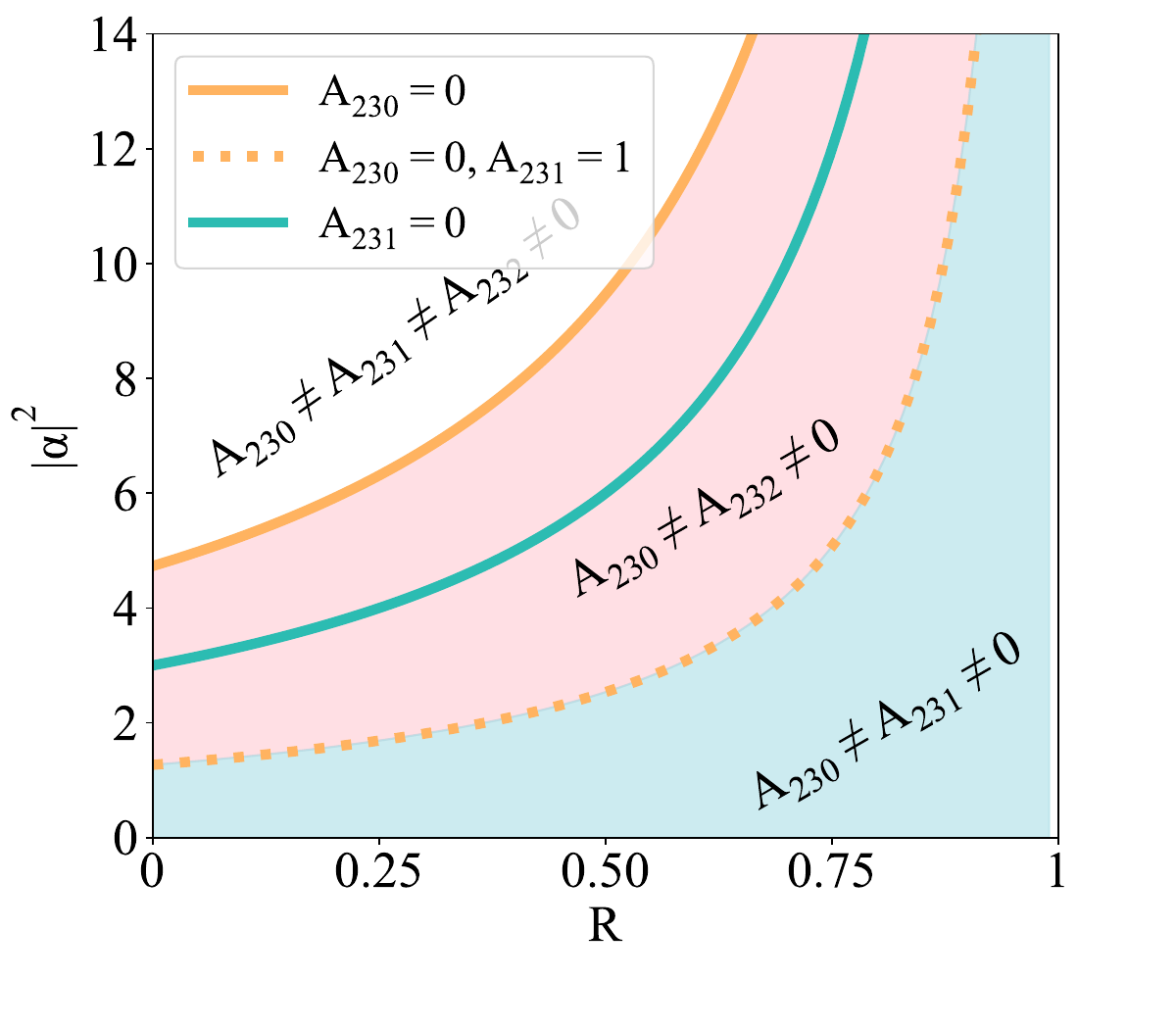}
         \subcaption[]{}
         \label{Fig.A23line}
     \end{subfigure}
     \caption{(a), (b), (c) and (d) display superposition coefficients in the parameter space $\abs{\alpha}^2$ and $R$ of the corresponding states $\hbox{DQ}_{2}^{+1}$, $\hbox{DQ}_{2}^{-1}$, $\hbox{DQ}_{3}^{+1}$ and $\hbox{DQ}_{3}^{-1}$ respectively.}
     \label{Fig.Scoeff}
\end{figure*}
In accordance with the detection of the photon state $m$, we classify the DQ as photon-added displaced qudits $(\hbox{DQ}^{+})$, photon-subtracted displaced qudits $(\hbox{DQ}^{-})$ and photon-catalyzed displaced qudits $(\hbox{DQ})$ states for $m<n$, $m>n$ and $m=n$, respectively. These representations are adapted from Dakna et al. \cite{Dakna1998b}. In this article, we focus mainly on differentiating $\hbox{DQ}^{+}$ and $\hbox{DQ}^{-}$ of CM for multiple state characterization and nonclassical properties. To simplify the wider spectrum of produced states, we refer to the states as follows:
\begin{itemize}
    \item If $m=n-2$, two-photon added displaced qudits $\hbox{DQ}^{+2}_{n}$.
    \item If $m=n-1$, single-photon added displaced qudits $\hbox{DQ}^{+1}_{n}$.
    \item If $m=n$, photon-catalyzed displaced qudits $\hbox{DQ}_{n}$.
    \item If $m=n+1$, single photon subtracted displaced qudits $\hbox{DQ}^{-1}_{n}$.
    \item If $m=n+2$, two-photon subtracted displaced qudits $\hbox{DQ}^{-2}_{n}$.
\end{itemize}
Moreover, the superposition coefficient of $\hbox{DQ}^{-}$ can be written in the simpler form \cite{Gorska2019},
\begin{align}
A_{nmq} = & N_{nm} \, \binom{n}{q} \, \sqrt{q!} \, \bigg( \frac{1-R}{R}\bigg)^{\frac{q}{2}} \, (-1)^{n-q} (n-q)! \nonumber \\ & \times
(\alpha^* \sqrt{1-R})^{m-n+q} \, \hbox{L}_{n-q}^{(m-n+q)} \big [ \abs{\alpha}^2 (1-R) \big].
\label{Eq.PADQ}
\end{align}
Here, $\hbox{L}_{m}^{(\alpha)}(\bullet)$ is the associated Laguerre polynomial. Similarly, the superposition coefficient of the $\hbox{DQ}^{+}$ state can be written as
\begin{align}
 A_{nmq} = & N_{nm} \, \binom{n}{q} \, \sqrt{q!} \, \bigg( \frac{1-R}{R}\bigg)^{\frac{q}{2}} \, (-1)^m m! \nonumber \\ & \times (\alpha \sqrt{1-R})^{n-m-q} \, \hbox{L}_{m}^{(n-m-q)} \big [\abs{\alpha}^2 (1-R) \big].
\label{Eq.PSDQ}
\end{align}
From the above equation, the order of the Laguerre polynomial is fixed to the detected photon state number $m$.  When we compared the superposition coefficients of $\hbox{DQ}^{+}$ with those of $\hbox{DQ}$ in \cite{Esakkimuthu2024}, it is observed that the coefficients are reduced to $1/(\alpha\sqrt{1-R})^y$, where $y$ is the number of photons added to the states. Meanwhile, for $\hbox{DQ}^{-}$ the coefficients are amplified to the rate of $(\alpha\sqrt{1-R})^z$ and $z$ represents the number of photons subtracted from the states. Hereafter, we will use the relation $\abs{\alpha}^2(1-R)=\chi$ for simplicity. This analysis explores the connections between $\chi$ and the superposition coefficients.

\subsection{Displaced qubits}
For the injection of a single photon state at the input port of CM and detection of any photon state at one output port, the resulting state at the other output port is called displaced qubits $(\hbox{DQ}_{2})$. 

\textbf{Case 1:} There is only one possibility of getting photon-added displaced qubits, $(\hbox{DQ}_{2}^{+1})$ when $m=n-1$. The single photon added displaced qudits $\hbox{DQ}_{2}^{+1}$ is computed as, 
\begin{align}
    \ket{\psi}_{10} = \hat{D}(\alpha\sqrt{R}) \, [A_{100} \ket{0} + A_{101} \ket{1}].
\end{align}
The coefficients of $\hbox{DQ}_{2}^{+1}$ are,
\begin{align*}
    A_{100} = N_{10} \, \alpha \sqrt{1-R}; \qquad
    A_{101} = N_{10} \, \sqrt{\frac{1-R}{R}}.
\end{align*}
The coefficients $A_{100}$ and $A_{101}$ express the nontrivial solutions of $\hbox{DQ}_{2}^{+1}$. However, we can find the equal superposition state $(1/\sqrt{2})[\ket{0}+\ket{1}]$ by fixing the coefficients. It comes under $\alpha=1/\sqrt{R}$ and it is shown in Fig. \ref{Fig.A10line} as a solid orange curve. The blue-filled regions in the figure constitute the greater possibility of getting the $\ket{1}$ state in $\hbox{DQ}_{2}^{+1}$. The remaining regions are dominated by the vacuum coefficient $A_{100}$.

\textbf{Case 2:}
A possibility of getting single photon subtracted displaced qubits, $(\hbox{DQ}_{2}^{-1})$ is when $m=n+1$. $\hbox{DQ}_{2}^{-1}$ is computed as, 
\begin{align}
    \ket{\psi}_{12} = \hat{D}(\alpha\sqrt{R}) \, [A_{120} \ket{0} + A_{121} \ket{1}].
\end{align}
The coefficients are,
\begin{align*}
    A_{120} = & N_{12} \, \alpha \sqrt{(1-R)} \, \big [\alpha^2 (1-R) - 2 \big]; \nonumber \\
    A_{121} = & N_{12} \, \alpha^2 \, \sqrt{\frac{(1-R)^3}{R}}.
\end{align*}
The strength of $A_{120}$ and $A_{121}$ in $\hbox{DQ}_{2}^{-1}$ are visualized in Fig. \ref{Fig.A12line} in the parameter space of $\alpha$ and $R$. The solid green curve in the figure represents the creation of a single-photon displaced state, $\hat{D}\big(\sqrt{R/(1-R)}\big) \ket{1}$. Naturally, the coefficient $A_{120}$ vanishes in this curve, i.e. $\chi (=2)$. Equal superposition states are generated at the output when $\alpha$ is equal to $(1/2\sqrt{R})[1\pm\sqrt{(1+7R)/(1-R)}]$. In the figure, the orange solid and dotted curves portray $(1/\sqrt{2})[\ket{0}+\ket{1}]$ and $(1/\sqrt{2})[\ket{0}-\ket{1}]$ for the sum and the difference in the above relation respectively. 
\subsection{Displaced qutrits}
The employment of two photons as an input state in the CM, the succeeded output state is referred to as the displaced qutrits $(\hbox{DQ}_{3})$. 

\textbf{Case 1:} Detecting a single photon in displaced qutrits prompts the creation of single photon added displaced qutrits $(\hbox{DQ}_{3}^{+1})$. From Eq. \ref{Eq.CMGen}, the state $\hbox{DQ}_{3}^{+1}$ is formulated as,
\begin{align}
    \ket{\psi}_{21} =  \hat{D}(\alpha\sqrt{R}) \, [A_{210} \ket{0} + A_{211} \ket{1} + A_{212} \ket{2}].
\end{align}
The coefficients of $\hbox{DQ}_{3}^{+1}$ is computed as,
\begin{align*}
    A_{210} = & \, N_{20} \, \alpha \sqrt{1-R} \, \big [\alpha^2 (1-R) - 2 \big] ;  \nonumber \\ 
    A_{211} = & N_{20} \, 2 \, \sqrt{\frac{1-R}{R}}  \, \big [\alpha^2 (1-R) - 1 \big]; \nonumber \\ 
    A_{212} = & \, N_{20} \, \sqrt{2} \, \alpha \frac{(1-R)^{\frac{3}{2}}}{R}.
\end{align*}
The variations of the coefficients $A_{210}$, $A_{211}$ and $A_{212}$ are illustrated in Fig. \ref{Fig.A21line}. The root of the coefficient $A_{210}$ is explicitly seen for $\chi = 2$, this condition is plotted as a solid orange line. In this circumstance, the state generates the superposition states $A_{211}\ket{1} + A_{212} \ket{2}$ with displacement. The coefficient $A_{211}$ also contains a root for the criteria $\chi$ equal to $1$ and it is visualized as a solid green line in Fig. \ref{Fig.A21line}. Here, the state is displaced superposition of $A_{210}\ket{0} + A_{212} \ket{2}$. Apparently, it demonstrates the formation of an even number state superposition with displacement. 

\textbf{Case 2:} The single photon-subtracted displaced qutrits $(\hbox{DQ}_{3}^{-1})$ is developed by the projection of three photons in $\hbox{DQ}_{3}$. It is expressed as
\begin{align}
    \ket{\psi}_{23} =  \hat{D}(\alpha\sqrt{R}) \, [A_{230} \ket{0} + A_{231} \ket{1} + A_{232} \ket{2}].
\end{align}
The coefficients are,
\begin{align*}
    A_{230} = & \, N_{23} \, \alpha \sqrt{1-R} \, \big [\alpha^4 (1-R)^2 - 6\alpha^2 (1-R) + 6 \big] ; \nonumber \\ 
    A_{231} = & \, N_{23} \, 2  \alpha^2 \, \sqrt{\frac{(1-R)^3}{R}}  \, \big [\alpha^2 (1-R) - 3 \big]; \nonumber \\ 
    A_{232} = &  N_{23} \, \sqrt{2} \, \alpha^3 \frac{(1-R)^{\frac{5}{2}}}{R}.
\end{align*}
Fig. \ref{Fig.A23line} displays the coefficient amplitudes $A_{230}$, $A_{231}$ and $A_{232}$ in the parameter space of $\alpha$ and $R$. The coefficient $A_{230}$ comprises two roots for the event of $\chi$ equals to $3 \pm \sqrt{3}$. The solid orange curve in the figure depicts the value $\chi$ belongs to $3+\sqrt{3}$, at this instance the state $\hbox{DQ}_{3}^{-1}$ is converted to $A_{231} \ket{1} + A_{232} \ket{2}$ with displacement. When $\chi=3-\sqrt{3}$, the coefficient of $\ket{0}$ vanishes and sequenced to the creation of $\hat{D}(1.1260\sqrt{R/(1-R)})\ket{1}$ with higher fidelity and it is represented as a dotted orange curve in the figure. The displacement in the present case is minimized compared to the displaced single photon generated in the $\hbox{DQ}_{2}$. The solid green curve in Fig. \ref{Fig.A23line} denotes the root of the coefficient $A_{231}$, which is positioned at $\chi = 3$ and manifests the generation of even superposition states $A_{230} \ket{0} + A_{232} \ket{2}$. 
\section{Quadrature Squeezing} \label{sec:QS}
\begin{figure*}[t]
     \begin{subfigure}{0.245\textwidth}
         \includegraphics[scale=0.22]{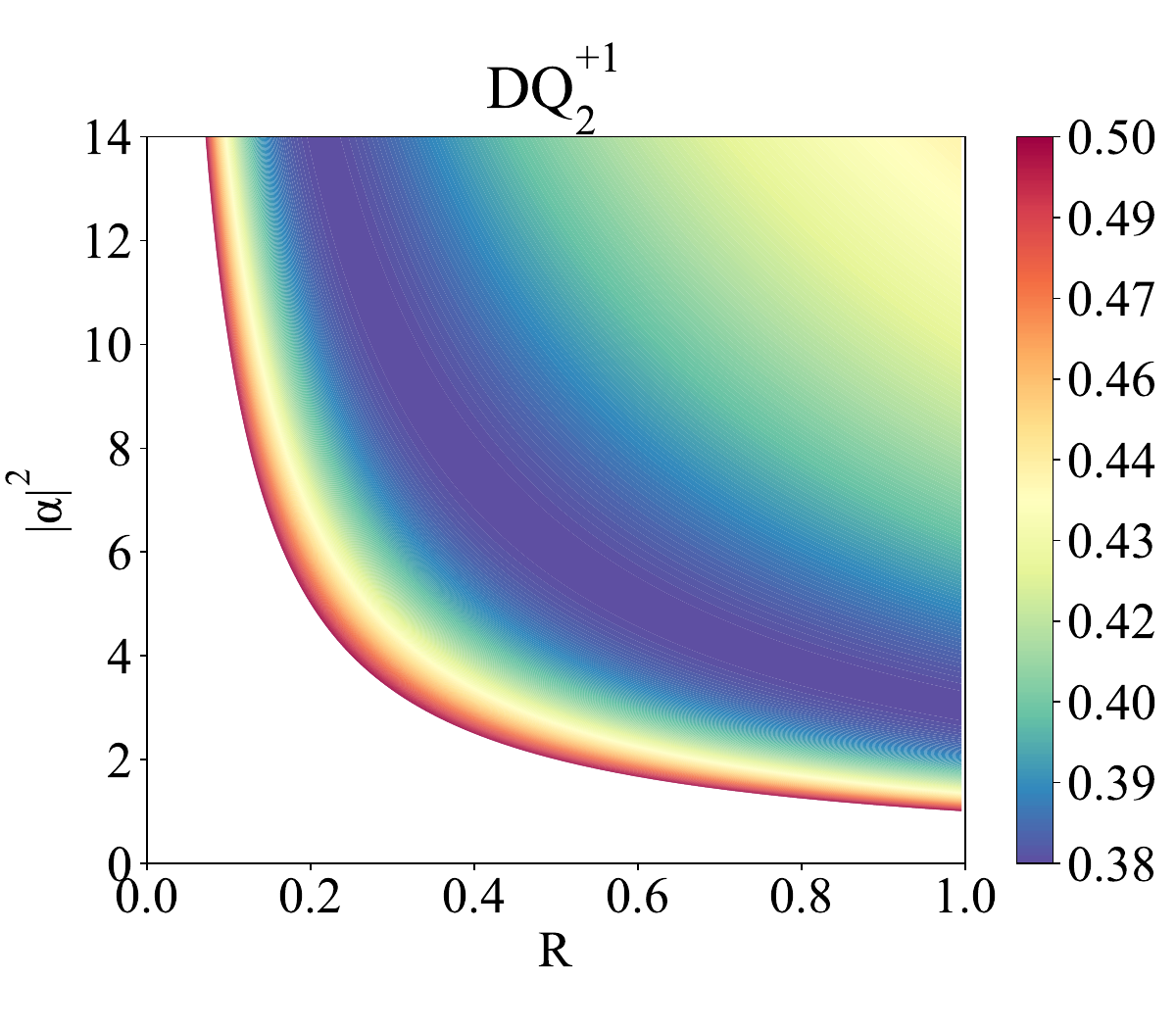}
         \subcaption[]{}
         \label{Fig.SQDQ20}
     \end{subfigure}
     \hfill
     \begin{subfigure}{0.245\textwidth}
         \includegraphics[scale=0.22]{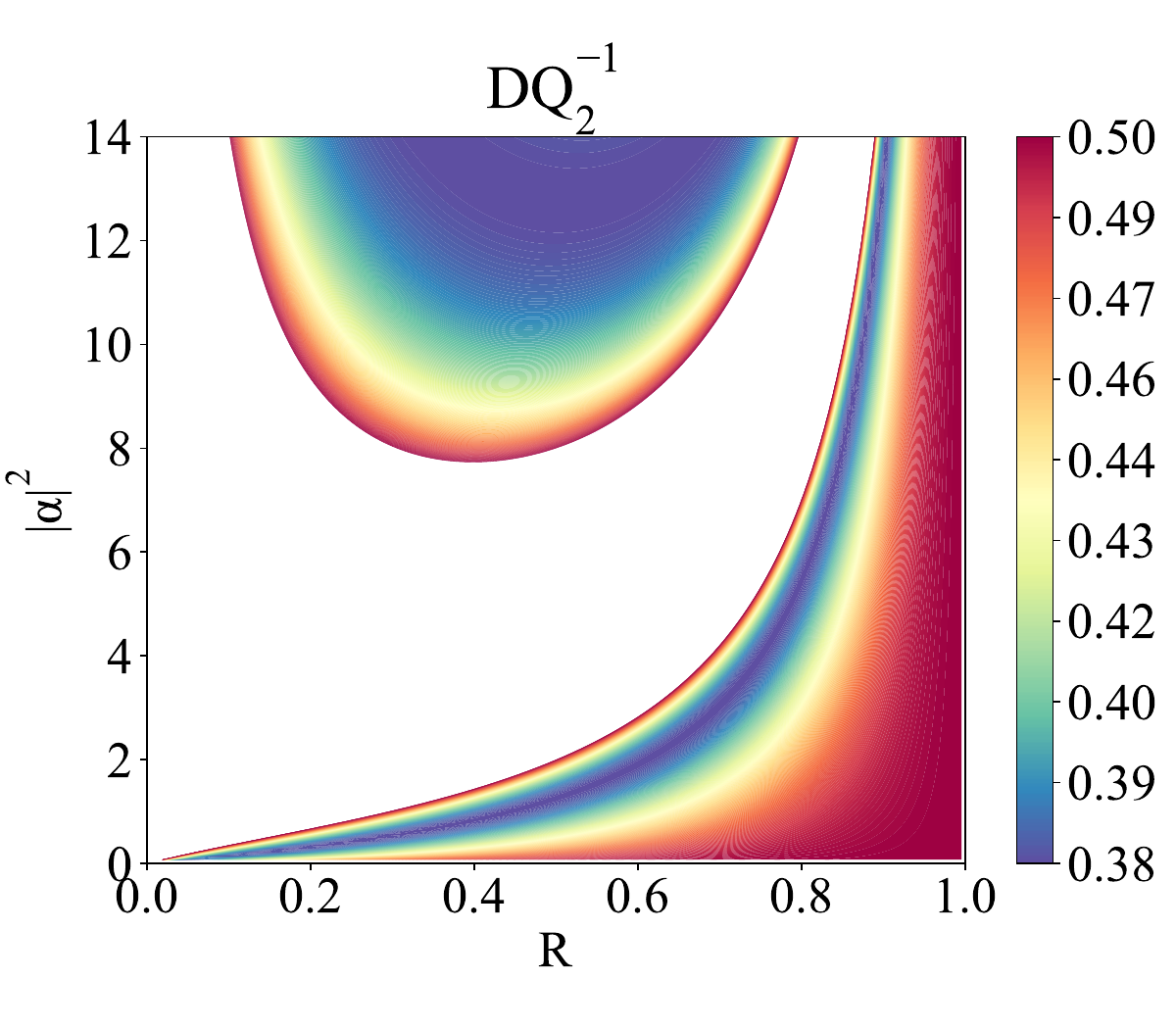}
         \subcaption[]{}
         \label{Fig.SQDQ22}
     \end{subfigure}
     \hfill
     \begin{subfigure}{0.245\textwidth}
         \includegraphics[scale=0.22]{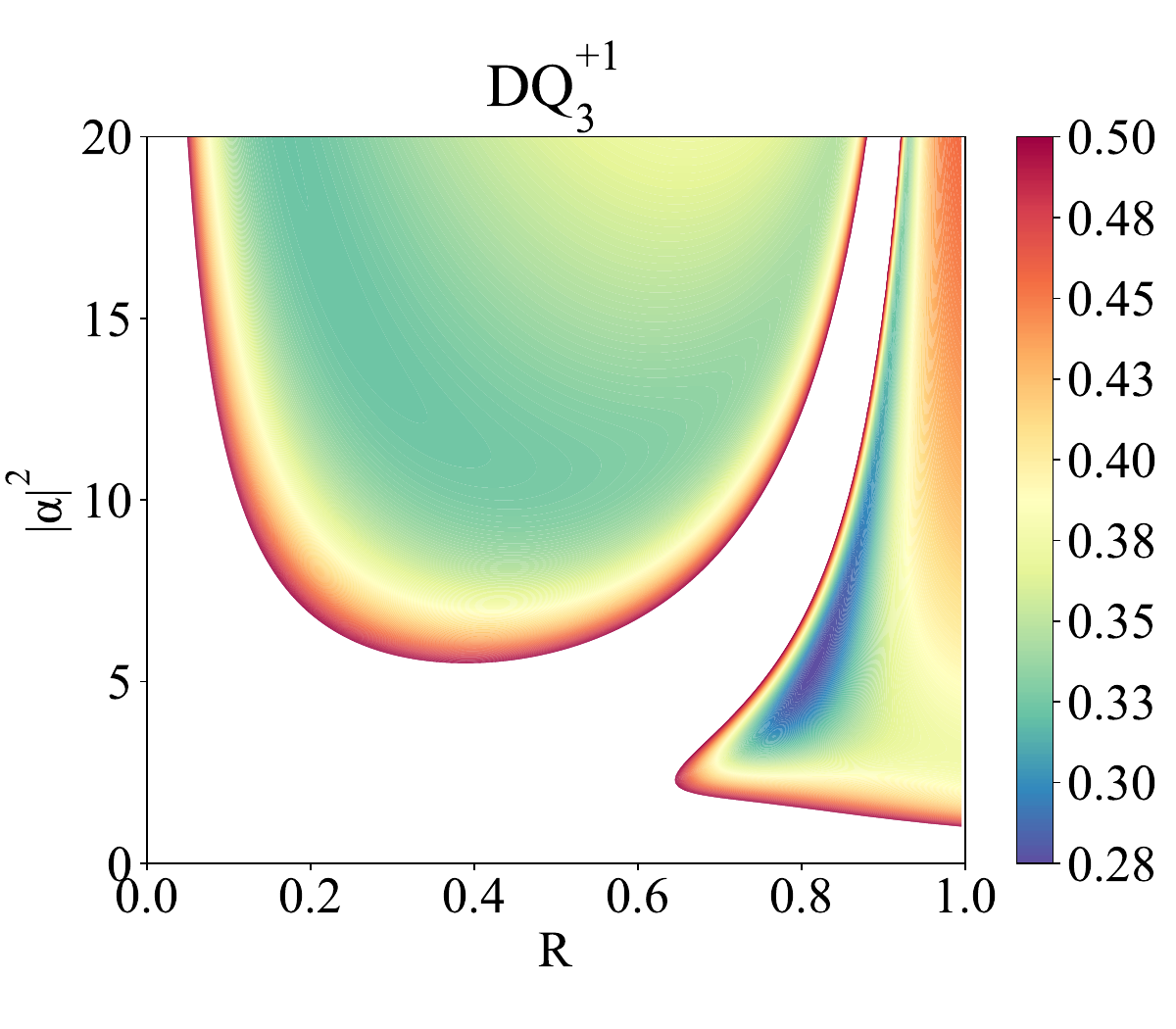}
         \subcaption[]{}
         \label{Fig.SQDQ31}
     \end{subfigure}
      \begin{subfigure}{0.245\textwidth}
         \includegraphics[scale=0.22]{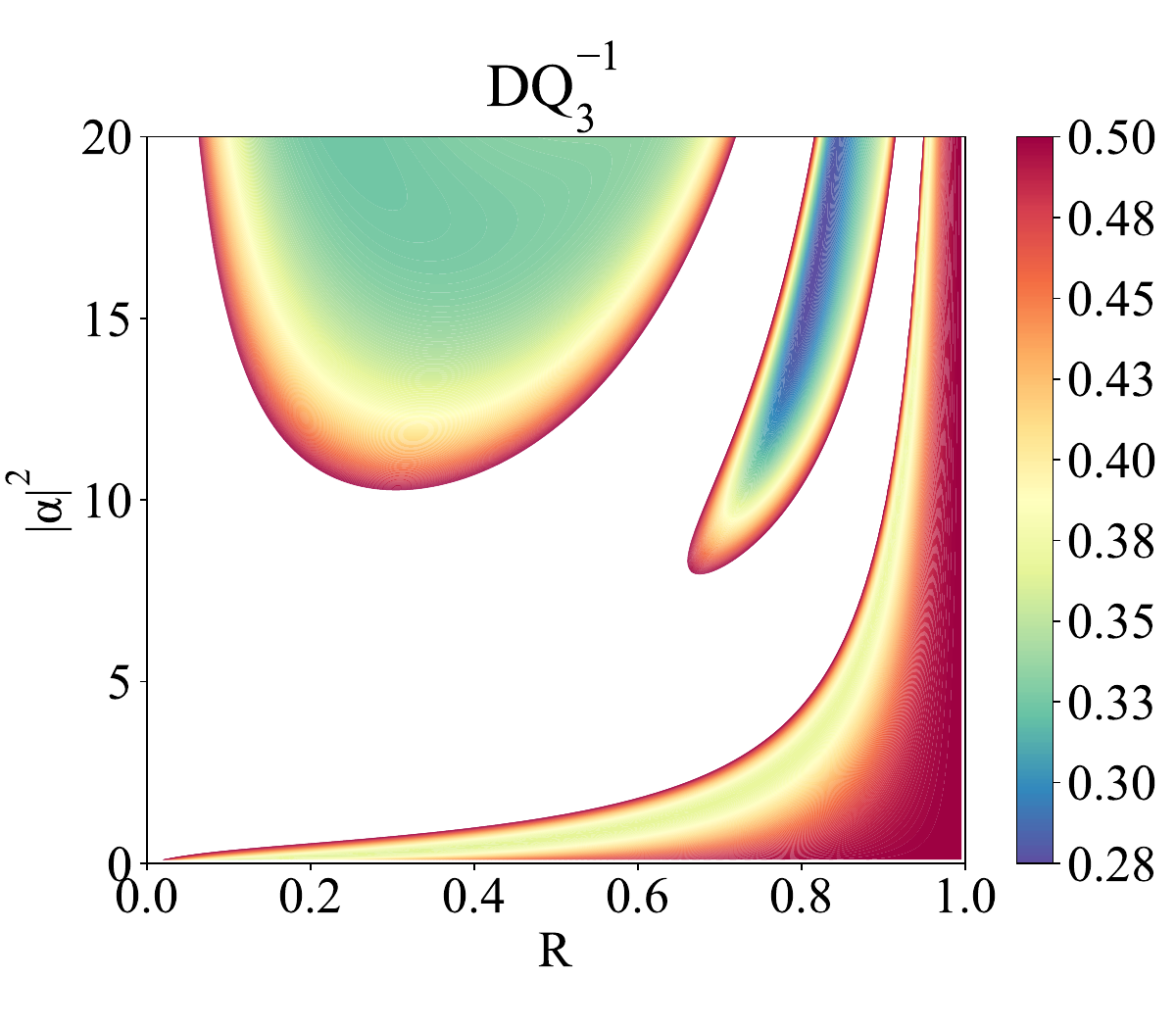}
         \subcaption[]{}
         \label{Fig.SQDQ33}
     \end{subfigure}
     \caption{(a), (b), (c) and (d) illustrate the quadrature squeezing of $\hbox{DQ}_{2}^{+1}$, $\hbox{DQ}_{2}^{-1}$, $\hbox{DQ}_{3}^{+1}$ and $\hbox{DQ}_{3}^{-1}$ respectively.}
     \label{Fig.SQDQ2}
\end{figure*}
The main goal of this article is to induce optimal squeezing features in the finite superposition of number states. Here, we turn to compute the quadrature squeezing properties of DQ to identify the achievable squeezing. Generally, the quadrature components of the optical field are considered as $\hat{X}= \frac{(\hat{a}+\hat{a}^{\dagger})}{\sqrt{2}}$ and $\hat{P}= \frac{(\hat{a}- \hat{a}^{\dagger})}{\sqrt{2}i}$. The corresponding quadrature variances can be computed by using the expressions, 
 \begin{align*}
    \expval{\Delta\hat{X}^2}= & \frac{1}{2} \Big[ \expval{\hat{a}^2}-\expval{\hat{a}}^2 + 
    \expval{\hat{a}^{\dagger 2}} - \expval{\hat{a}^{\dagger}}^2+ 2\expval{\hat{a}^{\dagger}\hat{a}} \\
    & - 2\expval{\hat{a}^{\dagger}}\expval{\hat{a}} + 1 \Big], \\
     \expval{\Delta\hat{P}^2}= &\frac{1}{2} \Big[ \expval{\hat{a}}^2 - \expval{\hat{a}^2} + \expval{\hat{a}^{\dagger}}^2 -\expval{\hat{a}^{\dagger 2}} + 2\expval{\hat{a}^{\dagger}\hat{a}} \\ & - 2\expval{\hat{a}^{\dagger}}\expval{\hat{a}} + 1 \Big].
\end{align*}
When any of the quadrature variances $\expval{\Delta\hat{X}^2}$ or $\expval{\Delta\hat{P}^2}$ is less than $1/2$, the state is considered squeezed. The minimum quadrature variance of DQ occurs in the $\hat{X}$ quadrature. The general form of higher-order moments of DQ is evaluated as,
\begin{align} 
    \expval{\hat{a}^{\dagger l}\hat{a}^{s}} = & \, \abs{\hbox{N}_{nm}}^2 \sum_{q=0}^{n} A_{nmq} \sum_{u=0}^{l} \binom{l}{u} \sum_{v=0}^s \binom{s}{v}  A_{nm(q-v+u)}^{*} \nonumber \\ & (\alpha^* \sqrt{R})^{l+s-u-v} \, \frac{[q!(q-v+u)!]^{1/2}}{(q-v)!}.
    \label{Eq.Mom}
\end{align}
Here, if the index of the coefficients possesses a negative value, it is considered as zero i.e., $(A_{nm(-ve)} = 0)$. The expressions of first and second-order moments can be obtained from Eq. \ref{Eq.Mom}, by substituting $l=0, s=1$ and $l=0, s=2$ respectively. 

Before discussing the squeezing features of DQ, two comments are in order. The first is the role of displacement operator and the second is the choice of superposition coefficients for optimal squeezing. The role of the displacement operator $\hat{D}(\beta)$ in the superposed states created was shown not to influence the squeezing of the state \cite{Lee1989}. This invariance arises from the fact that $\hat{D}(\beta)$ represents a Gaussian operation, which is incapable of manifesting the nonclassical properties of the system. In our second analysis, we compare the squeezing effects in DQ with the optimally achievable quadrature squeezing in Fock state superpositions ($\ket{\phi}_n$) \cite{Orlowski1995}.
\begin{align*}
        \ket{\phi}_n = \sum_{p=0}^{n} B_{np} \ket{p}
\end{align*}
The next step involves identifying the optimal amplitudes of the superposition coefficients in $\ket{\phi}_n$ to achieve maximal quadrature squeezing. For the state, $\ket{\Phi}_1 = B_{10}\ket{0}+B_{11}\ket{1}$ a maximum quadrature squeezing value of $0.3750$ is obtained when the coefficient amplitude satisfies $\abs{B_{10}}^2=3/4$. Furthermore, for state $\ket{\Phi}_2 = B_{20} \ket{0}+ B_{21}\ket{1}+B_{22}\ket{2}$, the minimum quadrature variance is calculated as $0.2753$ at the specific coefficient values $B_{20} = 0.9530, B_{21} = 0$ and $B_{22} = - 0.3030$.

We have compared the squeezing features of DQ in the parameter space of various $\alpha$ and $R$, see Fig. \ref{Fig.SQDQ2}. The dark blue regions in the figure indicate the maximum squeezing achieved in each case. Figs. \ref{Fig.SQDQ20} and \ref{Fig.SQDQ22} show the squeezing of $\hbox{DQ}_{2}^{+1}$ and $\hbox{DQ}_{2}^{-1}$, respectively. For both $\hbox{DQ}_{2}^{+1}$ and $\hbox{DQ}_{2}^{-1}$ states, the optimal squeezing observed is $0.3750$. A line range of squeezing is visible when $\alpha=\sqrt{{3}/{R}}$ where the superposition coefficients of $\hbox{DQ}_{2}^{+1}$ reach their required value of optimal squeezing, i.e. $\abs{A_{100}}^2 = 0.75$. For $\hbox{DQ}_{2}^{-1}$, optimal squeezing is achieved when $\abs{\alpha}^2$ equals $\dfrac{1}{4R}\left[\sqrt{3} \pm \sqrt{\frac{3+5R}{1-R}}\right]^2$. This expression also results in the coefficient $\abs{A_{120}}^2$ having a value of 0.75. 

Figs. \ref{Fig.SQDQ31} and \ref{Fig.SQDQ33} illustrate the squeeze properties of $\hbox{DQ}_{3}^{+1}$ and $\hbox{DQ}_{3}^{-1}$ individually. The maximum level of squeezing obtained from these states is $0.2753$. Compared to the optimal squeezing observed in the state $\ket{\Phi}_2$, one can also determine the optimal squeezing here, specifically when $\hbox{DQ}_{3}^{\pm}$ are defined to have roots of the coefficient $A_{2m1}$. Likewise for the $\hbox{DQ}_{3}^{+1}$ state, it manifests at the root of $\chi = 1$, which leads to the creation of $A_{210} \ket{0} + A_{212} \ket{2}$. In this instance, for a specific $\abs{\alpha}^2=5.45$ and $R=0.8175$, $\hbox{DQ}_{3}^{+1}$ attains the maximum squeezing. For the state $\hbox{DQ}_{3}^{-1}$, it takes place at $\chi=3$ to generate $A_{230} \ket{0} + A_{232} \ket{2}$. In the present case, the optimum squeezing is achieved at $\abs{\alpha}^2=10.95$ and $R=0.8175$. Comparing both $\hbox{DQ}_{3}^{\pm}$, the state $\hbox{DQ}_{3}^{+1}$ is observed to be the finest, as it produces optimal squeezing at the lower $\abs{\alpha}^2$.
\begin{table*}[htbp]  
\centering
\caption{\bf $\min (\Delta x)^2$ for respective input photon state $n$ and detected photon state $m$ in CM}
\begin{tabular}{cc@{}c@{}ccccccccc}
\hline
\multirow{2}{1em}{$m$} & \multicolumn{2}{c}{$n=1$} & \multicolumn{3}{c}{$n=2$} & \multicolumn{3}{c}{$n=3$} & \multicolumn{3}{c}{$n=4$} \\
\cline{2-12}
 & $\min (\Delta x)^2$ & $\abs{\alpha}^2$ & $\min (\Delta x)^2$ & $\abs{\alpha}^2$ & $R$ & $\min (\Delta x)^2$ & $\abs{\alpha}^2$ & $R$ & $\min (\Delta x)^2$ & $\abs{\alpha}^2$ & $R$ \\
\hline
$0$ & $0.3750$ & $\dfrac{3}{R}$ & $0.3223$ & $\dfrac{3}{R}$ & $\dfrac{3}{\abs{\alpha}^2}$ & $0.2913$ & $\dfrac{3}{R}$ & $\dfrac{3}{\abs{\alpha}^2}$ &  $0.2700$ & $\dfrac{3}{R}$ & $\dfrac{3}{\abs{\alpha}^2}$ \\
$1$ & $0.3750$ & $\dfrac{1}{4R}\bigg[\sqrt{3} \pm \sqrt{\frac{3+R}{1-R}} \bigg]^{2}$ & $0.2753$ & $5.45$ & $0.8175$ & $0.2353$ & $6.00$ & $0.7650$ & $0.2121$ & $6.65$ & $0.7275$ \\
$2$ & $0.3750$ & $\dfrac{1}{4R}\bigg[\sqrt{3} \pm \sqrt{\frac{3+5R}{1-R}}\bigg]^{2}$ & $0.2753$ & $10.95$ & $0.8175$ & $0.2448$ & $13.75$ & $0.7975$ & $0.2288$ & $16.65$ & $0.7875$ \\
$3$ & $0.3750$ & $\dfrac{1}{4R}\bigg[\sqrt{3} \pm \sqrt{\frac{3+9R}{1-R}}\bigg]^{2}$ & $0.2753$ & $16.45$ & $0.8175$ & $0.2489$ & $21.60$ & $0.8125$ & $0.2411$ & $14.50$ & $0.8600$ \\
$4$ & $0.3750$ & $\dfrac{1}{4R}\bigg[\sqrt{3} \pm \sqrt{\frac{3+11R}{1-R}}\bigg]^{2}$ & $0.2753$ & $21.63$ & $0.8175$ & $0.2511$ & $28.87$ & $0.8175$ & $0.2447$ & $23.02$ & $0.8675$ \\
\hline
\end{tabular}
  \label{tab:Compare}
\end{table*}

Table \ref{tab:Compare} provides a comprehensive overview of the achievable quadrature squeezing throughout the configuration of the CM setup. When considering a single-photon input state $(n=1)$, the maximum squeezing is achieved over a distribution of any photon detection and holds a certain relation between $\abs{\alpha}^2$ and $R$ which is denoted in the table. Moreover, for vacuum detection conditions $(m=0)$ the maximal squeezing is achieved under the condition of $\abs{\alpha}^2 = 3/R$ for arbitrary input photon states. For inputs $n>2$, the optimal squeezing of DQ is obtained by detecting a single-photon state $(m=1)$ for the corresponding $\abs{\alpha}^2$ and $R$. This is due to the fact that to obtain the maximal squeezing, the amplitude of $A_{nm0}$ must attain higher values relative to the other coefficients. Under the detection condition $m=1$, the superposition coefficient $A_{n10}$, shows a single root, for the relation of $\chi=n$. Thus, single-photon detection event facilitates maximal squeezing compared with the other events. 

Table. \ref{tab:m1} shows the difference between the attainable squeezing of CM and the optimal squeezing obtained from $\ket{\phi}_n$. The reduction in the peak values observed for the input states $n>2$ arises from the limited ability to adjust the superposition coefficients effectively. For example, optimal squeezing of the $n=1,2$ states is achieved by generating superpositions of two number states, where the desired superposition amplitudes can be easily adjusted under conditions $\abs{A_{1m0}}^2+\abs{A_{1m1}}^2=1$ and $\abs{A_{2m0}}^2+\abs{A_{2m2}}^2=1$. However, for the input states $n>2$, all superposition coefficients must be nonzero to achieve optimal squeezing, requiring a greater number of adjustable parameters.

\begin{table}[htbp]  
\centering
\caption{\bf min$(\Delta x)^2$ obtained in $\ket{\phi}_{n}$ \cite{Orlowski1995} and in DQ at the $m=1$ detection}
\begin{tabular}{ccccc|}
\hline
$n$ & $\hbox{DQ}_{n+1}^{+(n-1)}$ & $\ket{\phi}_n$ & Difference \\
\hline
$1$ & $0.3750$ & $0.3750$ & $0$ \\
$2$ & $0.2753$ & $0.2753$ & $0$ \\
$3$ & $0.2353$ & $0.2298$ & $0.0055$ \\
$4$ & $0.2121$ & $0.1902$ & $0.0219$ \\
$5$ & $0.1963$ & $0.1645$ & $0.0318$ \\
$6$ & $0.1845$ & $0.1451$ & $0.0394$ \\
\hline
\end{tabular}
  \label{tab:m1}

\end{table}
\section{Non-Gaussianity} \label{sec:NG}
\begin{figure*}[t]
     \begin{subfigure}{0.245\textwidth}
         \includegraphics[scale=0.22]{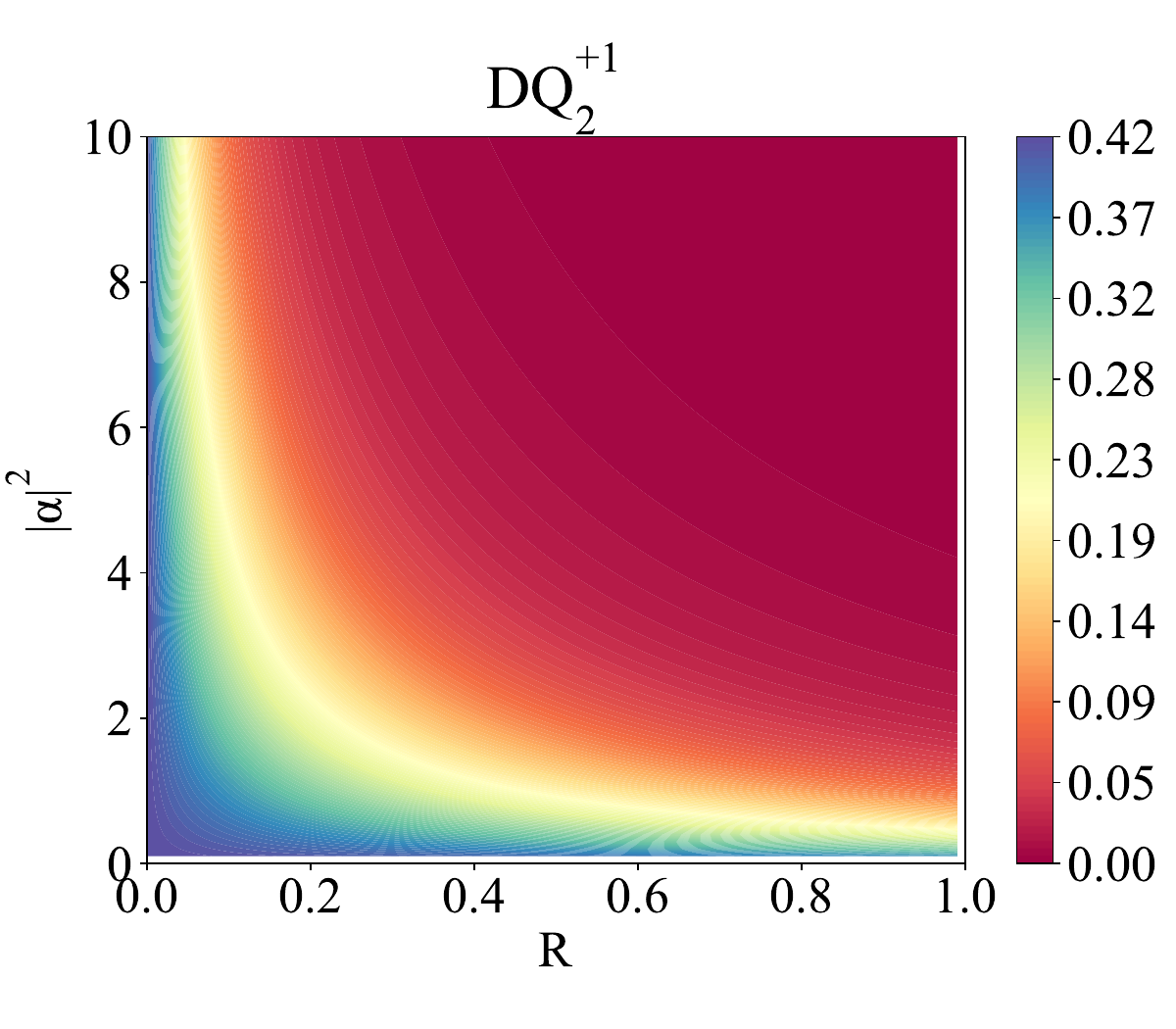}
         \subcaption[]{}
         \label{Fig.HSDDQ20}
     \end{subfigure}
     \hfill
     \begin{subfigure}{0.245\textwidth}
         \includegraphics[scale=0.22]{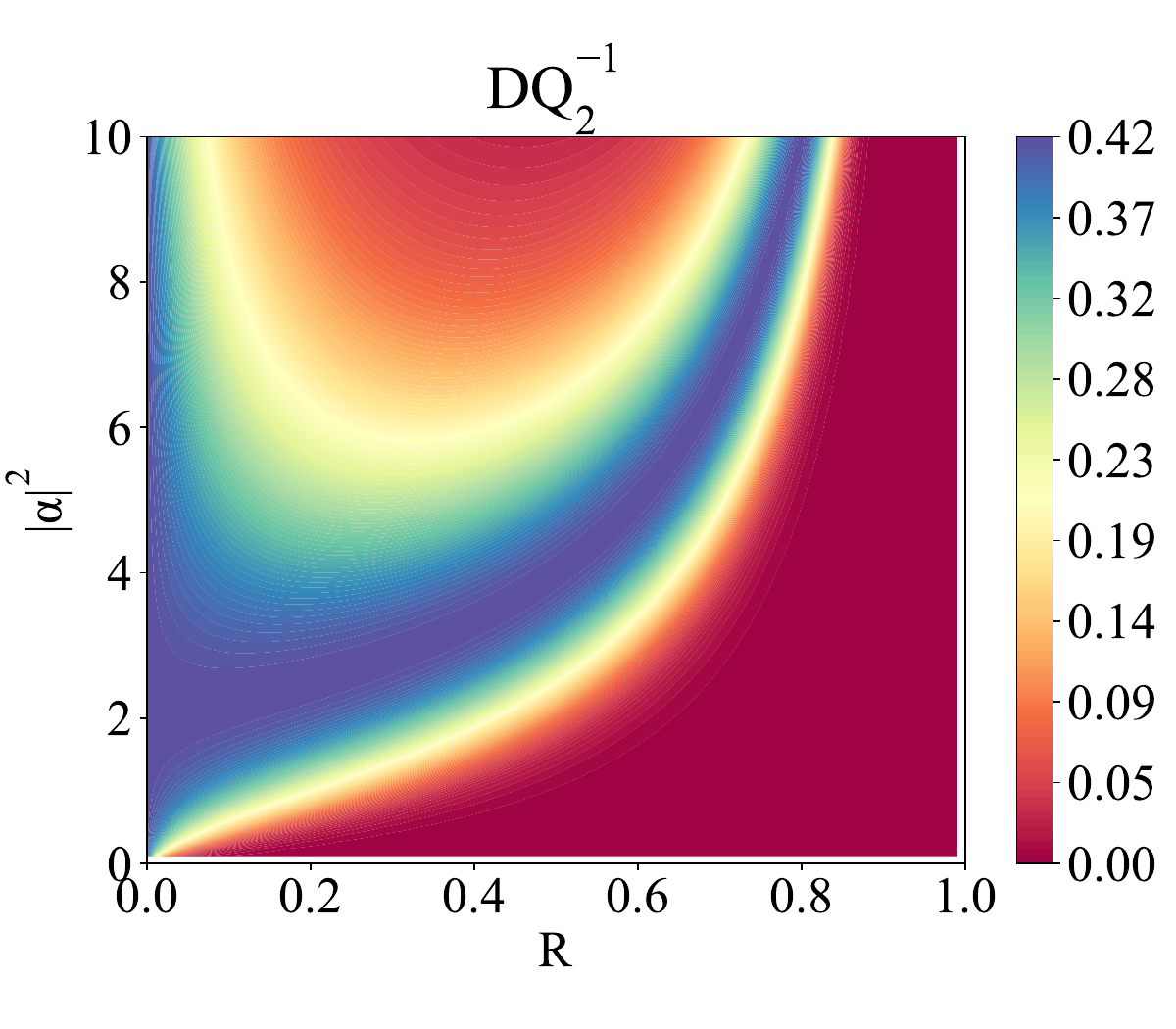}
         \subcaption[]{}
         \label{Fig.HSDDQ22}
     \end{subfigure}
     \hfill
     \begin{subfigure}{0.245\textwidth}
         \includegraphics[scale=0.22]{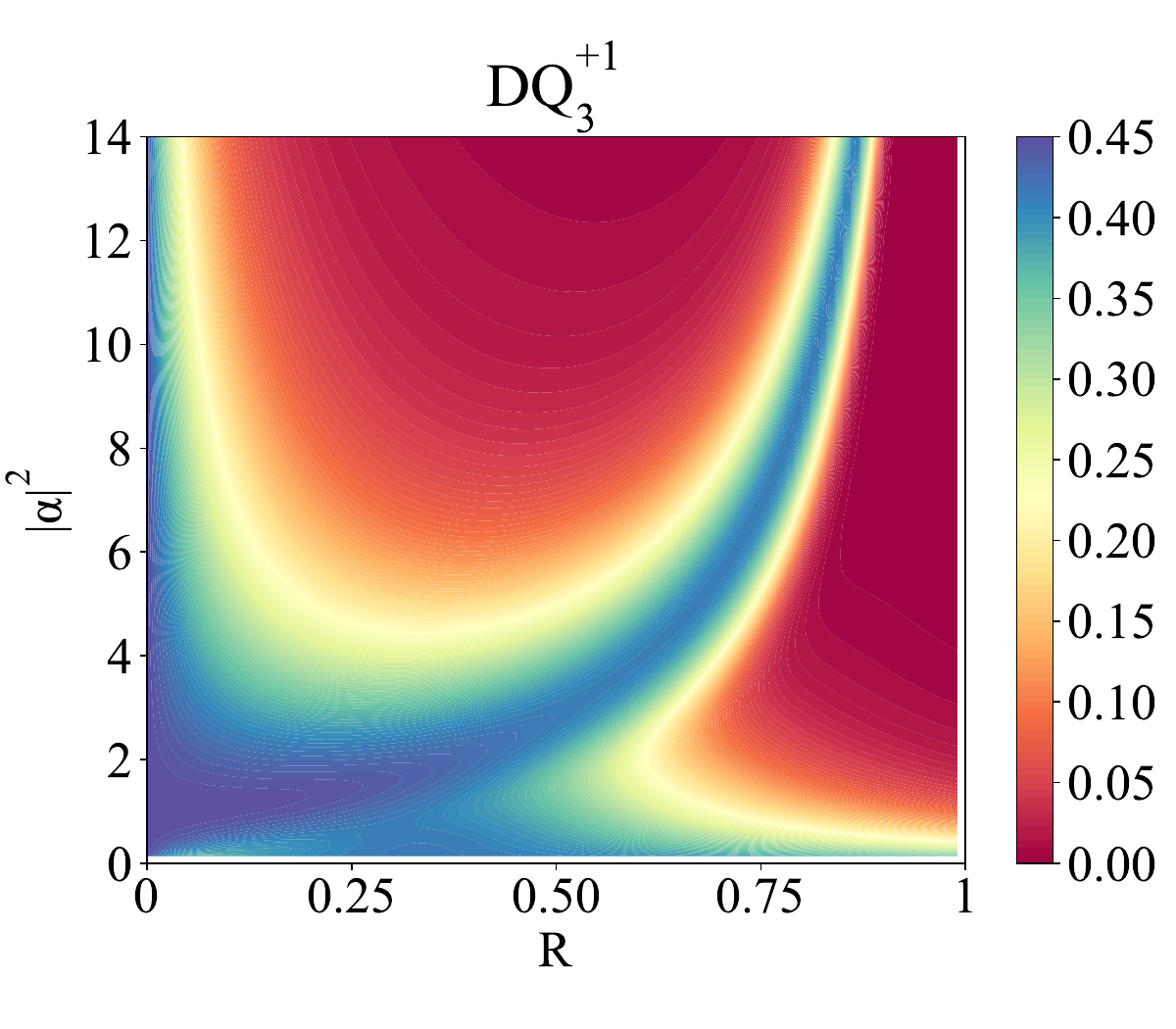}
         \subcaption[]{}
         \label{Fig.HSDDQ31}
     \end{subfigure}
      \begin{subfigure}{0.245\textwidth}
         \includegraphics[scale=0.22]{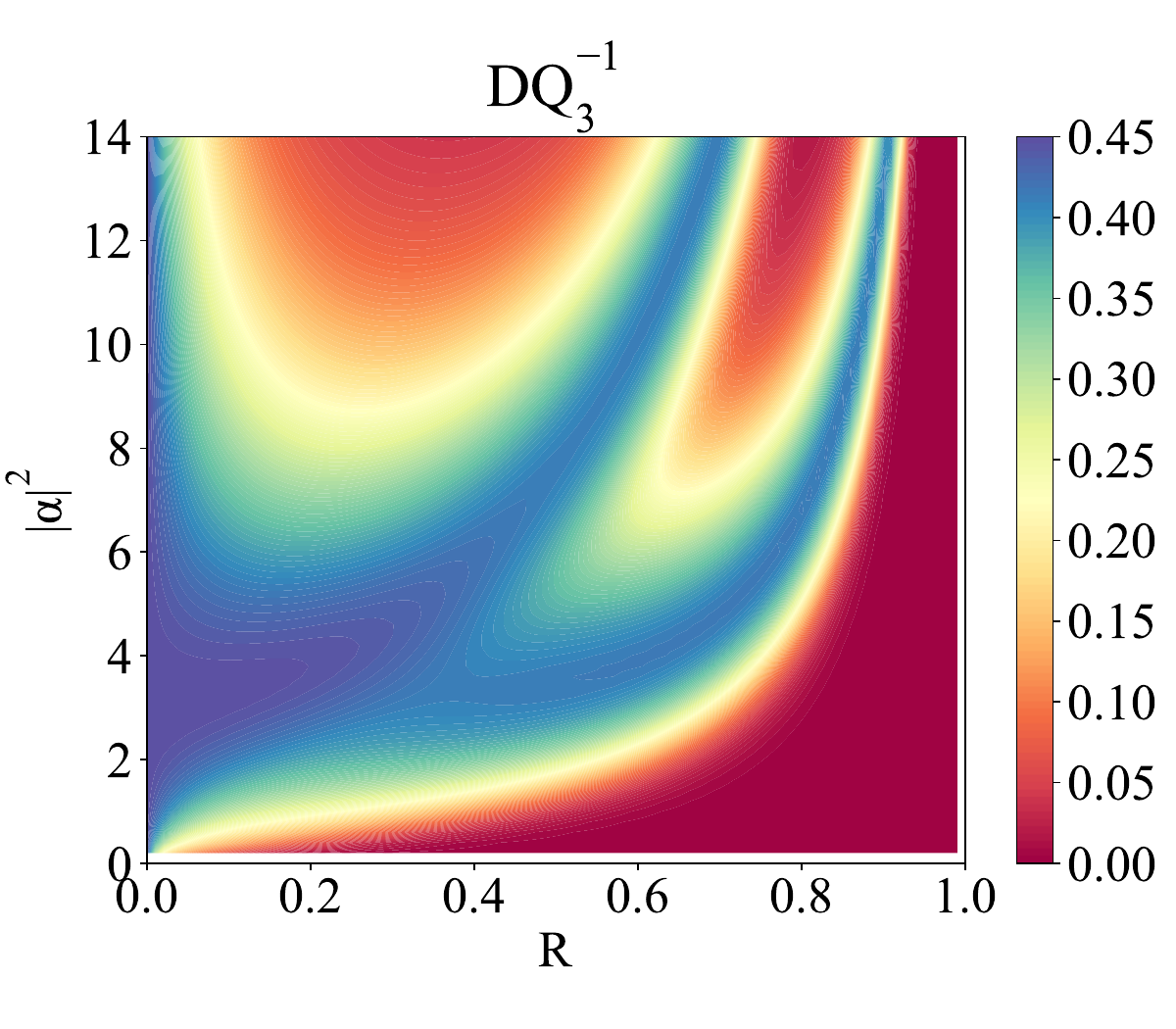}
         \subcaption[]{}
         \label{Fig.HSDDQ33}
     \end{subfigure}
     \caption{(a), (b), (c) and (d) display the HSD values of $\hbox{DQ}_{2}^{+1}$, $\hbox{DQ}_{2}^{-1}$, $\hbox{DQ}_{3}^{+1}$ and $\hbox{DQ}_{3}^{-1}$ respectively in the parameter space of $\abs{\alpha}^2$ and $R$.}
     \label{Fig.HSDDQ2}
\end{figure*}
In this part, we assess the amount of non-Gaussianity exhibited by the generated output states. Evidence of non-Gaussian characteristics by a state can be quantified using various non-Gaussian metrics \cite{Paris2007, Paris2008, Xiang2018, Chabaud2020}. In particular, we use two measures to characterize the non-Gaussian nature of DQ namely the Hilbert-Schmidt distance and the Wigner negativity.
\subsection{Hilbert Schmidt Distance}
Genoni et al. \cite{Paris2007} proposed a measure to evaluate the non-Gaussianity present in the state based on the Hilbert-Schmidt distance (HSD). This measure involves the estimation of the geometric distance between the quantum state ($\rho$) under investigation and a reference Gaussian state ($\tau$). The HSD of the given $\rho$ can be evaluated by, 
\begin{equation*}
\delta[\rho] = \frac{\hbox{Tr}[(\rho - \tau)^2]}{2\hbox{Tr}(\rho^2)}.
\end{equation*}
When $\delta[\rho]>0$, the state can be identified as a non-Gaussian state. The upper bound of HSD measure is $\frac{1}{2}$. In this investigation, we have taken the displaced squeezed thermal state as a reference state \cite{Satya1985, Marian1993}. The density operator of this reference state is described by,
\begin{equation*}
\tau =\hat{D}(\gamma)\hat{S}(\zeta)\nu(\bar{n})\hat{D}^{\dagger}(\gamma)\hat{S}^{\dagger}(\zeta)
\end{equation*}
where, $\hat{D}(\gamma)=e^{\gamma\hat{a}^\dagger-\gamma^*\hat{a}}$ is the displacement operator and $\hat{S}(\zeta)= e^{1/2[\zeta\hat{a}^{\dagger 2}-\zeta^*\hat{a}^2]}$, $(\zeta=r e^{i\phi})$ represents the squeezing operator.
\begin{align*}
\nu(\bar{n}) =  \,  \left(1+\bar{n}\right)^{-1}\Big[ \frac{\bar{n}}{1+\bar{n}}\Big]^{a^{\dagger}a} =  \, \sum_{k=0}^{\infty} \frac{\bar{n}^k}{\left(1+\bar{n}\right)^{k+1}}\ket{k}\bra{k}, 
\end{align*}
$\nu(\bar{n})$ signifies the thermal state, where $\bar{n}$ denotes the average number of photons. By comparing the first and second moments of $\rho$ and $\tau$, we can calculate the parameters of the reference state ($\gamma, \gamma^*,\phi, r, \bar{n}$). 

The HSD values of $\hbox{DQ}_{2}^{+1}$ and $\hbox{DQ}_{2}^{-1}$ are illustrated in Fig. \ref{Fig.HSDDQ20} and \ref{Fig.HSDDQ22} respectively. The maximum non-Gaussianity observed for both cases is $0.4167$. The state $\hbox{DQ}_{2}^{-1}$ has more non-Gaussian regions comparing with the $\hbox{DQ}_{2}^{+1}$. The dark blue patches in the figure are the non-Gaussian of $\hat{D} \big(\sqrt{R/(1-R)}\big)\ket{1}$ and in this event the optimal non-Gaussianity is observed. Besides the maximal value, the non-Gaussianity of $\hbox{DQ}_{2}^{+1}$ and $\hbox{DQ}_{2}^{-1}$ are visible in the other generation of superposition state, where the coefficients of $\ket{1}$ are more prominent than the coefficient of $\ket{0}$. Removing the vacuum state in the superposed states asserts the non-Gaussianity of $\hbox{DQ}_{2}$. 
The HSD values of $\hbox{DQ}_{3}^{+1}$ and $\hbox{DQ}_{3}^{-1}$ are illustrated in Fig. \ref{Fig.HSDDQ31} and \ref{Fig.HSDDQ33} separately. The maximum non-Gaussianity for both cases is observed as $0.4518$. For $\hbox{DQ}_{3}^{+1}$,the non-Gaussianity is enhanced in the blue regions for the value $\chi$ of $2$ which leads to the superposed state $A_{211} \ket{1} + A_{212} \ket{2}$, see Fig. \ref{Fig.HSDDQ31}. In $\hbox{DQ}_{3}^{-1}$ the optimal HSD values are noticed as two sub-dark lines. From lower to higher $R$ the first and second dark blue regions are due to the existence of $A_{231}\ket{1}+A_{232}\ket{2}$ and $A_{231}\ket{1}$ with displacement respectively. By linking the figures \ref{Fig.HSDDQ2} of HSD and \ref{Fig.SQDQ2} of quadrature squeezing, one can predict the complementary presence of non-Gaussianity and quadrature squeezing within the states. We can see that the regions with zero values of HSD have nonzero quadrature squeezing. This behavior was previously discussed in the Appendix A of \cite{Gerry2018}. Furthermore, it is observed that the CM's output non-Gaussianity does not improve than the input non-Gaussianity. To enhance non-Gaussianity, more nonclassical resources are required.
\subsection{Wigner distribution}
The Wigner distribution offers a qualitative approach to evaluate the nonclassicality and non-Gaussianity of the quantum state \cite{Kenfack2004,Genoni2013}. In the coherent basis, the Wigner function is given by,
\begin{equation*}
W(\beta,\beta^*)=\frac{2}{\pi^2}\int d^2 z \bra{-z}\hat{\rho}\ket{z}e^{-2(z\beta^* -z^*\beta )}.
\end{equation*}
$\hat{\rho}$ represents the density operator of the state examined. The negative volume of the Wigner function quantifies the state's non-Gaussianity. The Wigner negativity $(W_{{N}})$ volume can be computed by using,
\begin{equation*}
     W_{{N}} = \int \int \abs{W(\beta,\beta^*)} \, d\beta \,d\beta^*  -1.
\end{equation*}
The Wigner function of DQ is derived as,
\begin{align*}
W(\beta) = & \, \abs{\hbox{N}_{nm}}^2 e^{2\abs{\beta}^2-\abs{A}^2} \frac{2}{\pi} \sum_{p,q=0}^{n} \frac{A_{nmp}A_{nmq}^{*}}{\sqrt{p!}\sqrt{q!}}  \\ & \sum_{u=0}^{p} \binom{p}{u} (\alpha^*\sqrt{R})^{p-u} \sum_{v=0}^{q} \binom{q}{v} (\alpha \sqrt{R})^{q-v} 
\hbox{H}_{u,v} (A^*,A).
\end{align*}
Here, $A=\alpha\sqrt{R}-2\beta$ and $\hbox{H}_{n,m}(X^*,X)$ is the two-variable Hermite polynomial. 
\begin{figure*}[t]
     \begin{subfigure}{0.245\textwidth}
         \includegraphics[scale=0.22]{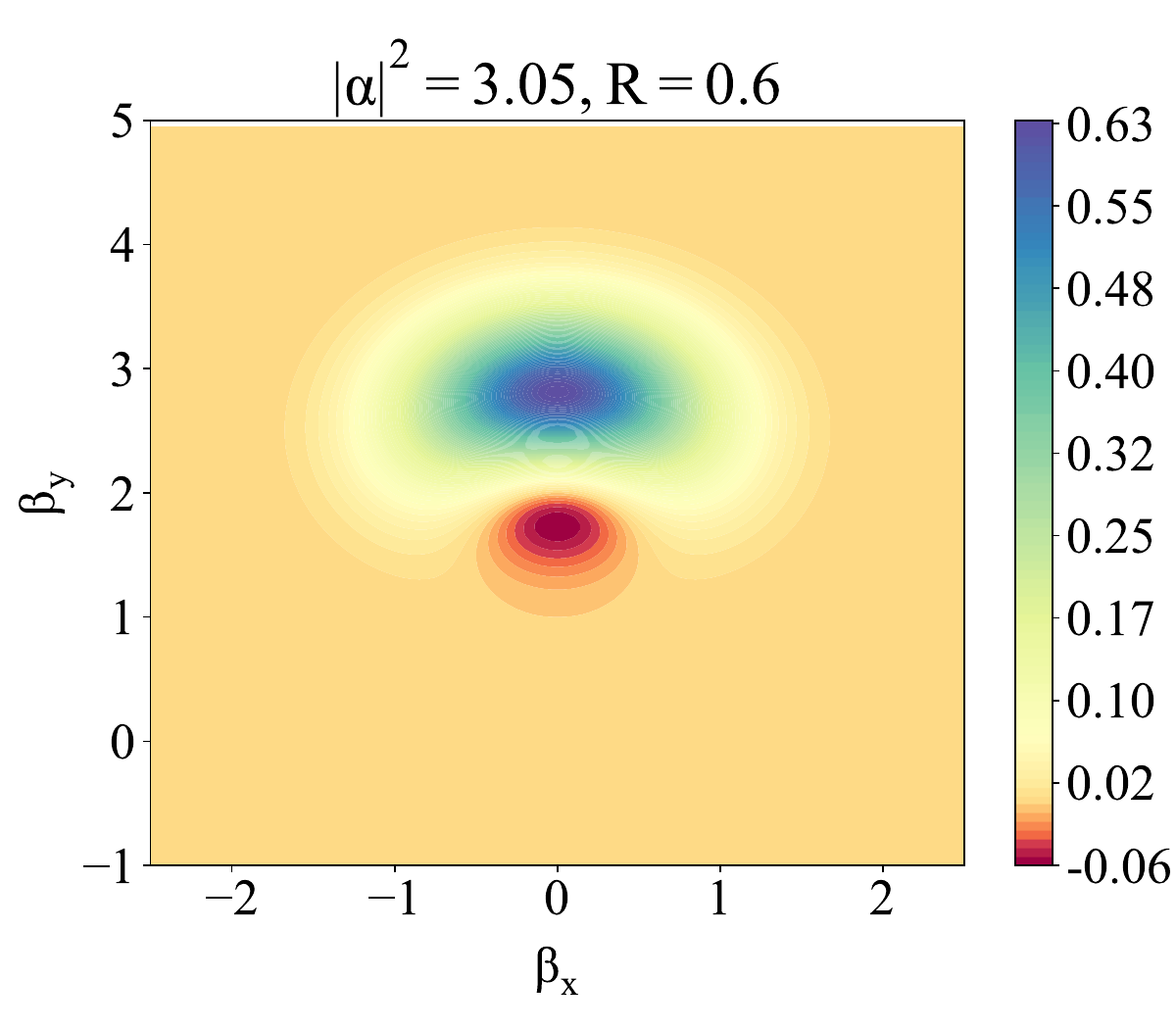}
         \subcaption[]{}
         \label{Fig.Wign1}
     \end{subfigure}
     \hfill
     \begin{subfigure}{0.245\textwidth}
         \includegraphics[scale=0.22]{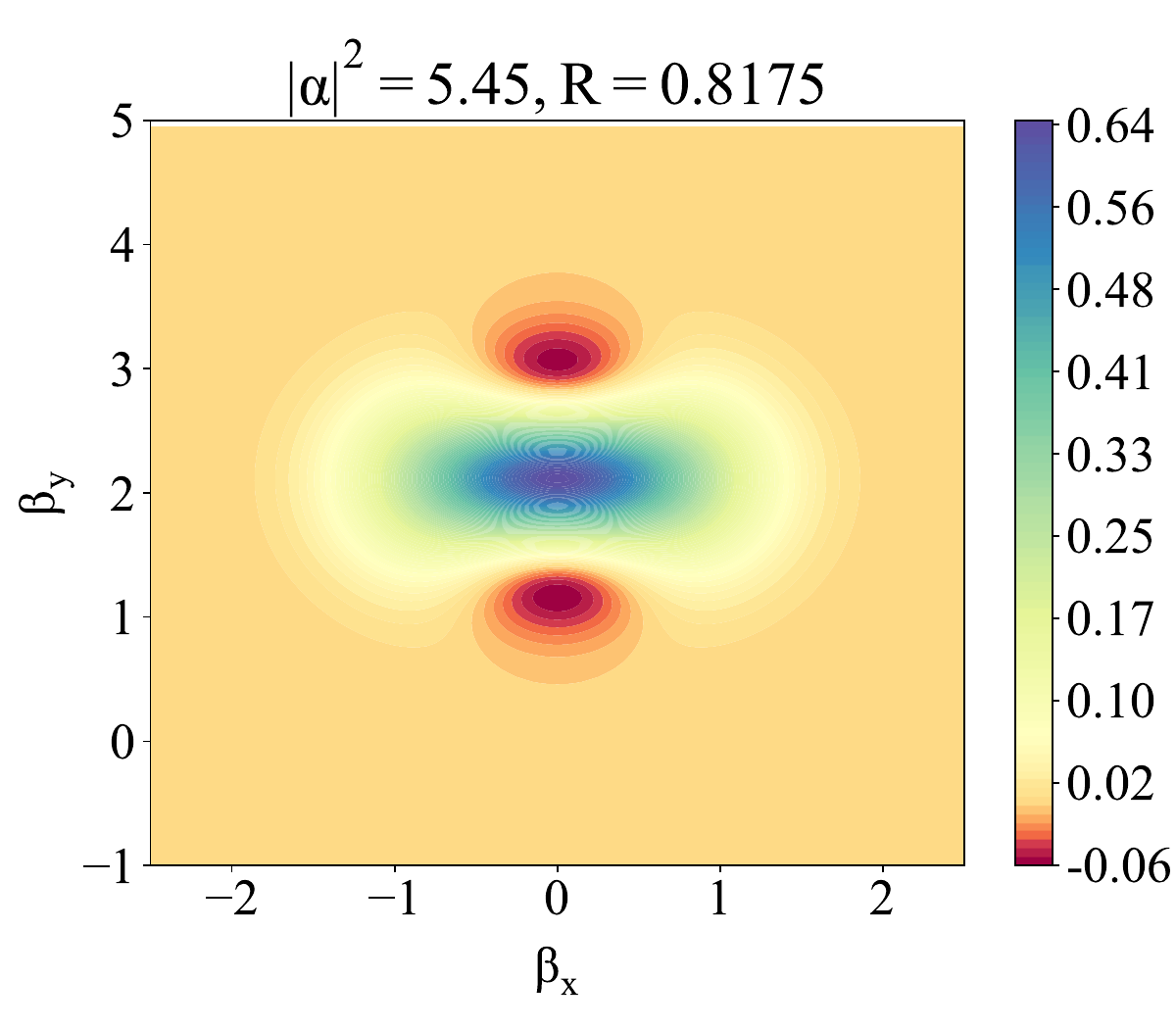}
         \subcaption[]{}
         \label{Fig.Wign2}
     \end{subfigure}
     \hfill
     \begin{subfigure}{0.245\textwidth}
         \includegraphics[scale=0.22]{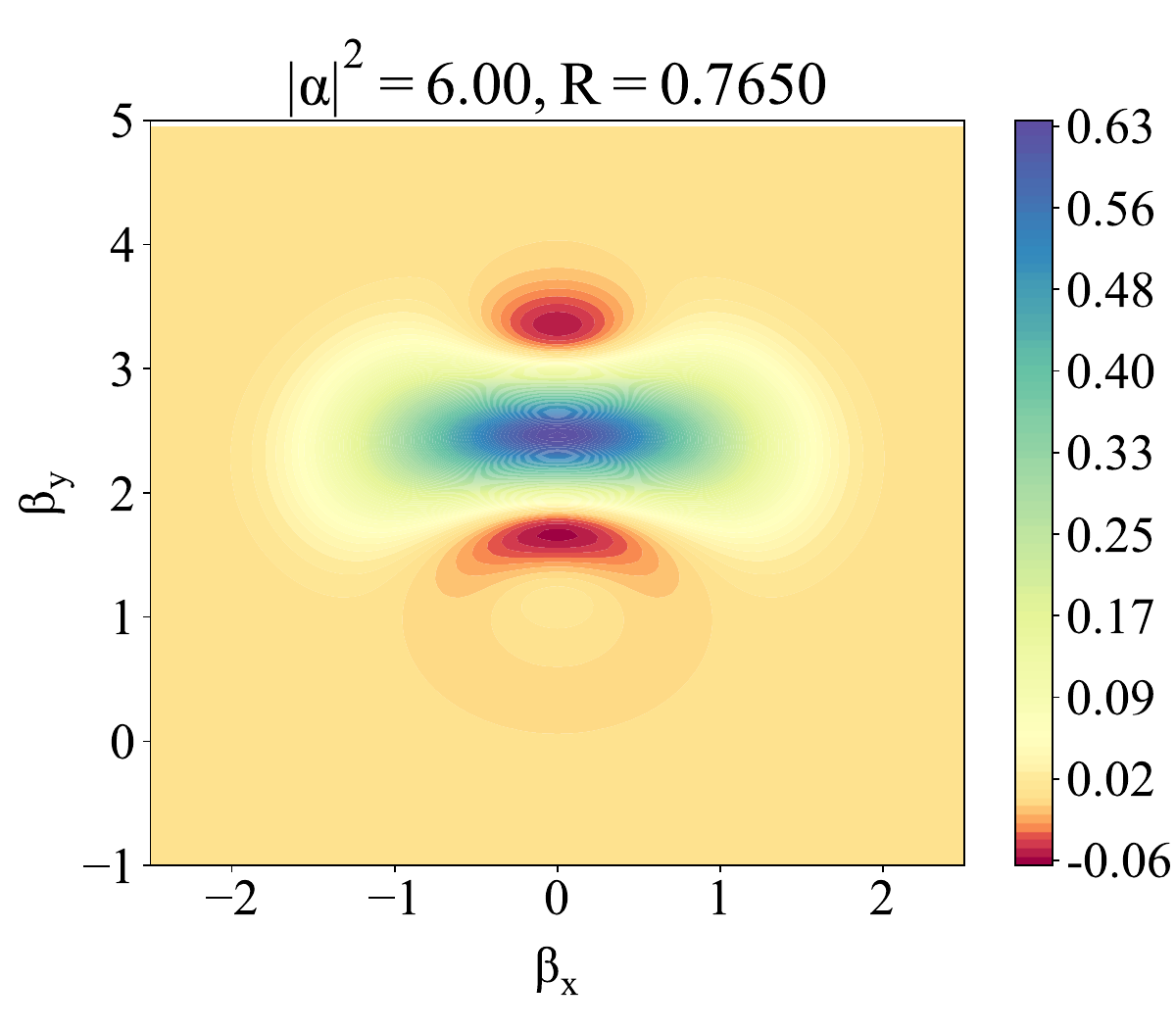}
         \subcaption[]{}
         \label{Fig.Wign3}
     \end{subfigure}
      \begin{subfigure}{0.245\textwidth}
         \includegraphics[scale=0.22]{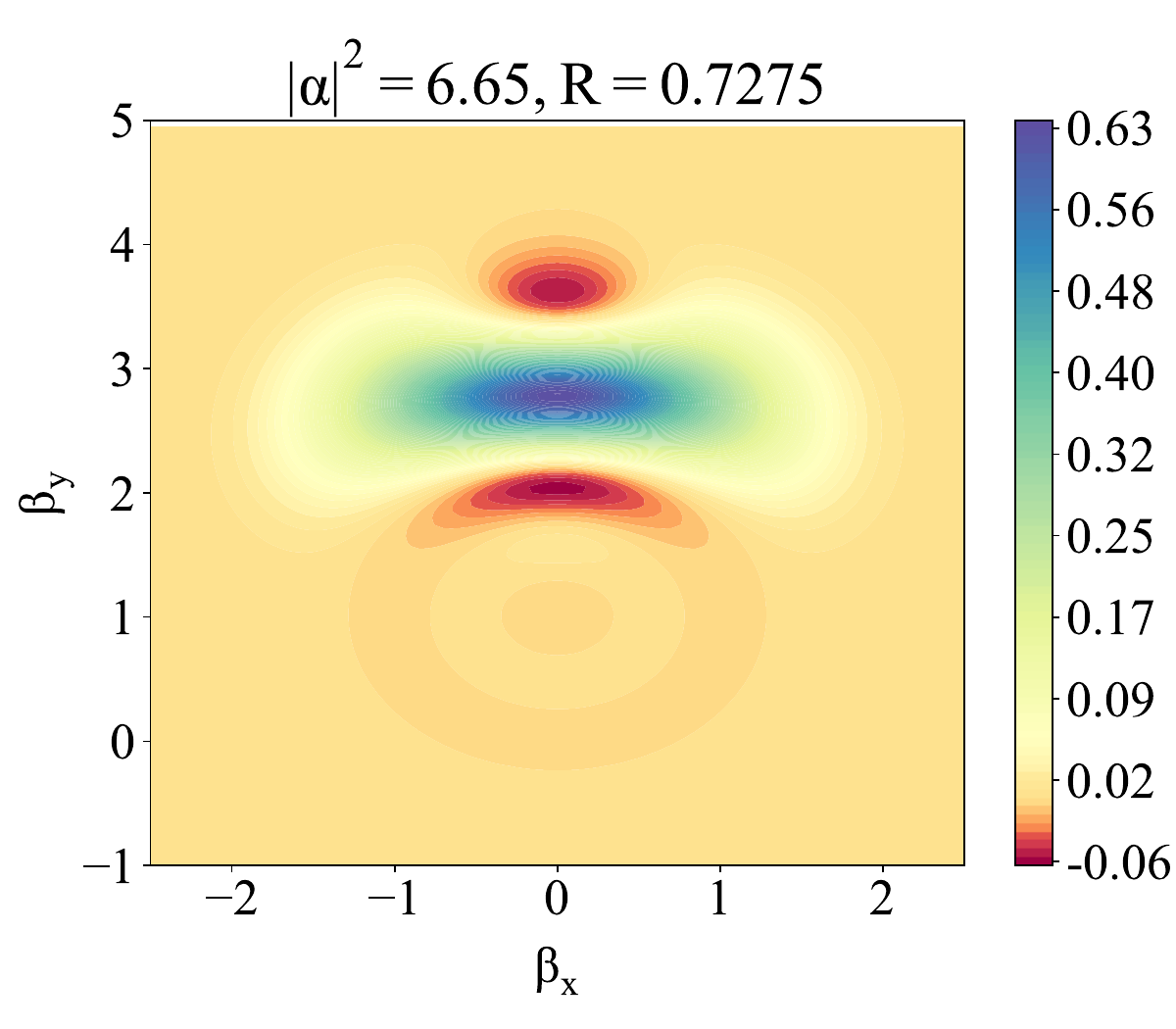}
         \subcaption[]{}
         \label{Fig.Wign4}
     \end{subfigure}
     \caption{(a), (b), (c) and (d) present the Wigner distribution of the optimal squeezed states achieved from $\hbox{DQ}_{2}$, $\hbox{DQ}_{3}^{+1}$, $\hbox{DQ}_{4}^{+2}$ and $\hbox{DQ}_{5}^{+3}$ individually.}
     \label{Fig.Wig}
\end{figure*}
The Wigner distribution of the respective states with optimal squeezing for each input state is shown in Fig. \ref{Fig.Wig}. 
Fig. \ref{Fig.Wign1} presents the Wigner distribution of $\hbox{DQ}_{2}$ for $m=1$, where a single stretched hump is observed, which clearly demonstrating the squeezed nature of the state. Figures. \ref{Fig.Wign2}, \ref{Fig.Wign3} and \ref{Fig.Wign4} portrays the Wigner distribution of $\hbox{DQ}_{3}^{+1}$, $\hbox{DQ}_{4}^{+2}$ and $\hbox{DQ}_{5}^{+3}$ separately. Altogether, these figures reveal an increasing degree of stretching (squeezing) with respect to the increase the photon number of the input number states, as visible in the blue regions. The Wigner negativity for the corresponding Wigner distributions is summarized in the table. \ref{tab:SP}. We observe a slight increase in the Wigner negativity as the input number states increase. This trend can be analyzed by examining Fig. \ref{Fig.Wig}, where negative regions become more pronounced with higher input number states.

\section{Non-ideal experimental conditions} \label{sec:NI}

In this section, we examine the imperfections in the experimental realization of DQ, focusing on the practical impact of photon detector inefficiency and the mixed-state nature of photon sources. We adopted the same experimental realization scheme methodology presented in \cite{Esakkimuthu2024}. Without dark counts, the POVM for a photon-number-resolving detector identifying an $l$ photon state is given by \cite{Scully1969},
\begin{equation}
    \hat{\pi}_l = \sum_{k=l}^{\infty} \binom{k}{l} \eta_d^{l} (1-\eta_d)^{k-l} \ket{k} \bra{k}.
    \label{Eq.DI}
\end{equation}
Here, $\eta_d$ denotes the efficiency of the heralding detector, with $(0 \leq \eta_d \leq 1)$. The input photon's mixed nature is expressed as a convex combination of the vacuum state and photon number states $(n)$ \cite{Bimbard2010},
\begin{equation}
    \hat{\rho}_a = (1-\eta_s) \ket{0}\bra{0} + \eta_s \ket{n}\bra{n},
\end{equation}
where $\eta_s$ denotes the non-ideal efficiency of the photon sources. These imperfections alter the pure state of Eq. \ref{Eq.CMGen} into a mixed state. Thus, the input state becomes the density operator $\hat{\rho}_{in} = \ket{\alpha}\bra{\alpha} \otimes \hat{\rho}_a$. After the BS unitary operation and the effects of the imperfect detector, the resulting non-ideal DQ state is:
\begin{align}
    \hat{\rho}_{R} = & \sum_{k=m}^{\infty} \binom{k}{m} \eta_d^{m} (1-\eta_d)^{k-m} \bra{k} \hat{U}^{\dagger}_{BS} \hat{\rho}_{in} \hat{U}_{BS} \ket{k}.
    \label{Eq.RQOC}
\end{align}
\subsection{Fidelity}
\begin{figure*}[t]
     \begin{subfigure}{0.245\textwidth}
         \includegraphics[scale=0.22]{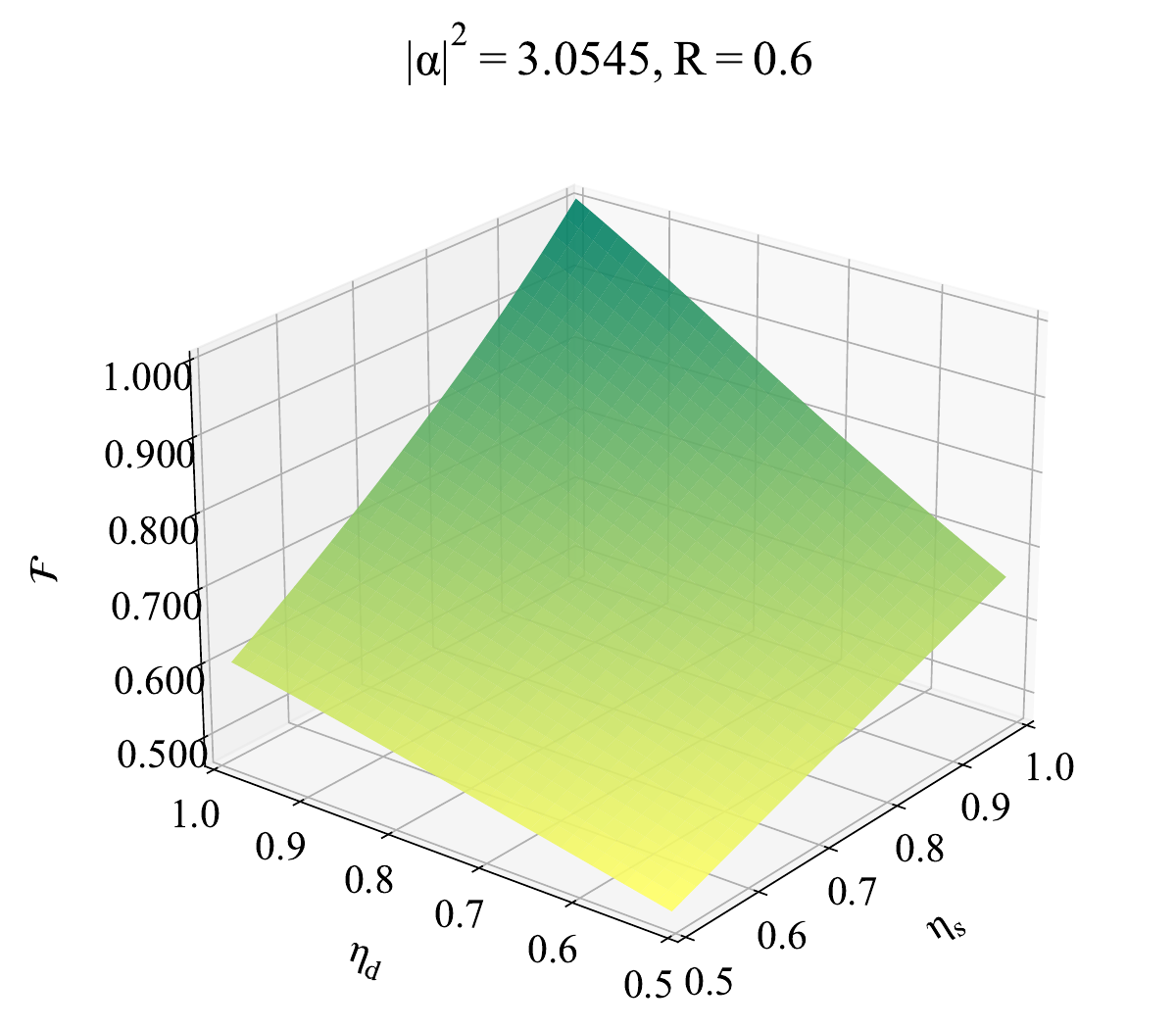}
         \subcaption[]{}
         \label{Fig.FidDQ21}
     \end{subfigure}
     \hfill
     \begin{subfigure}{0.245\textwidth}
         \includegraphics[scale=0.22]{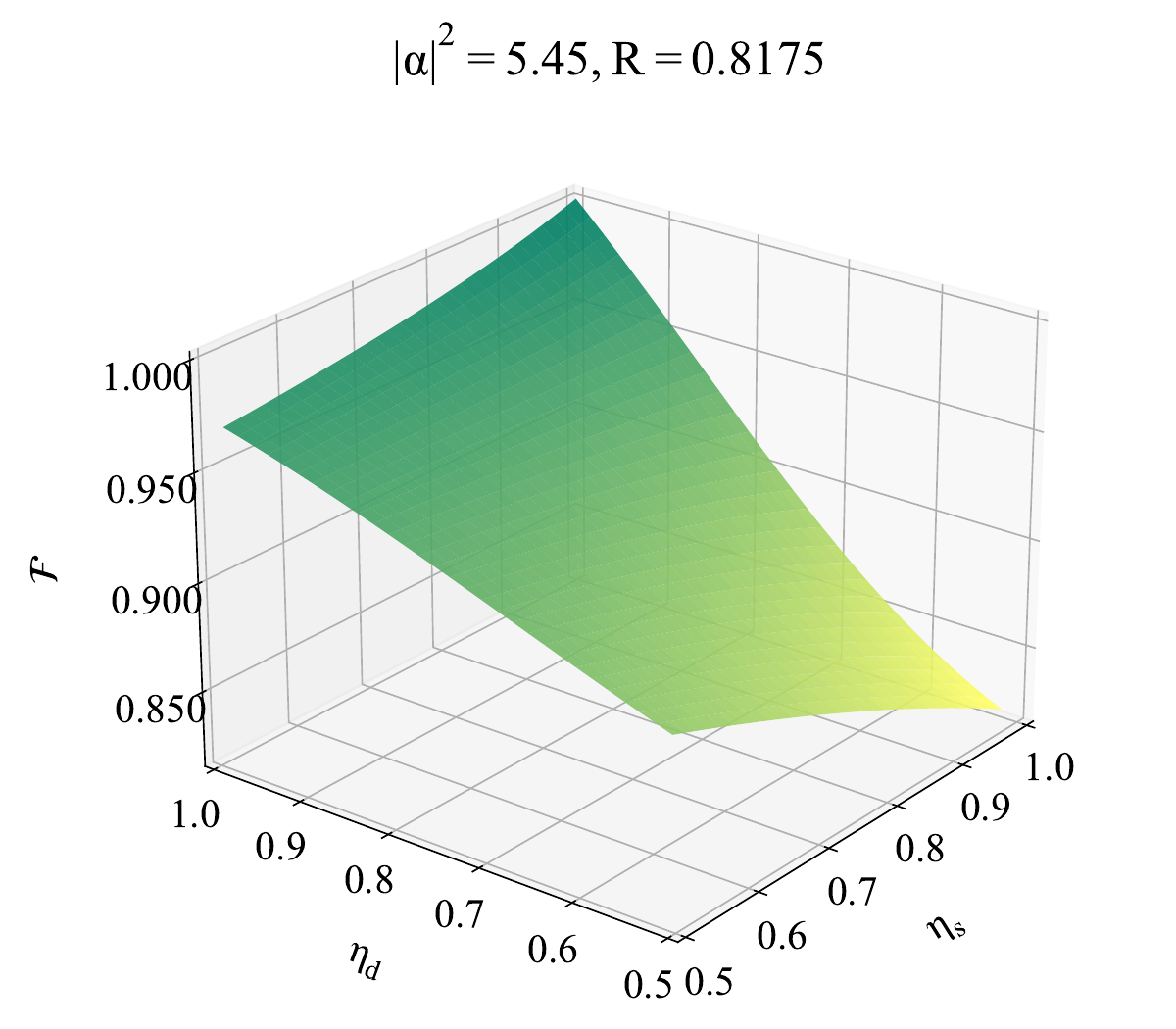}
         \subcaption[]{}
         \label{Fig.FidDQ31}
     \end{subfigure}
     \hfill
     \begin{subfigure}{0.245\textwidth}
         \includegraphics[scale=0.22]{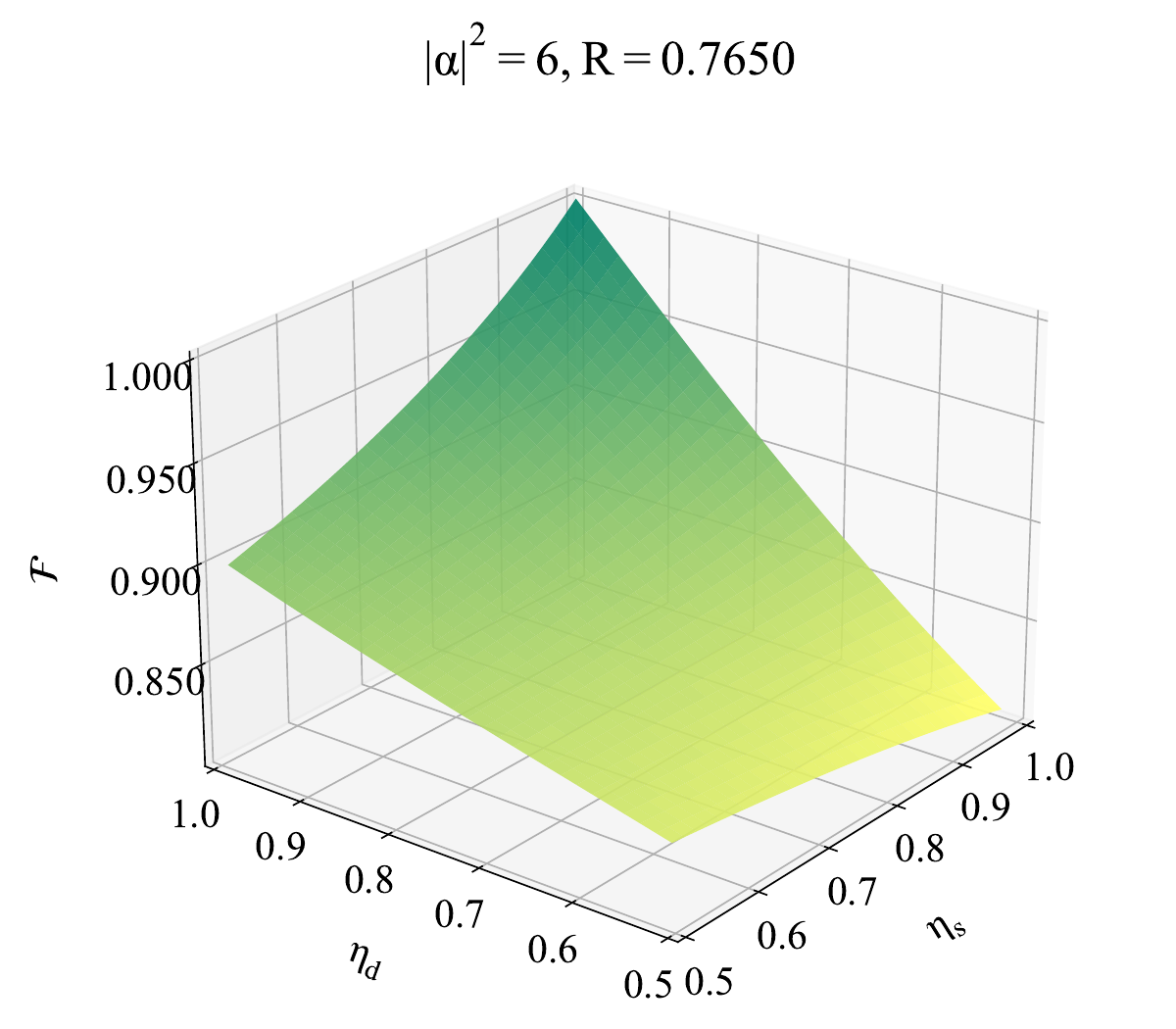}
         \subcaption[]{}
         \label{Fig.FidDQ41}
     \end{subfigure}
      \begin{subfigure}{0.245\textwidth}
         \includegraphics[scale=0.22]{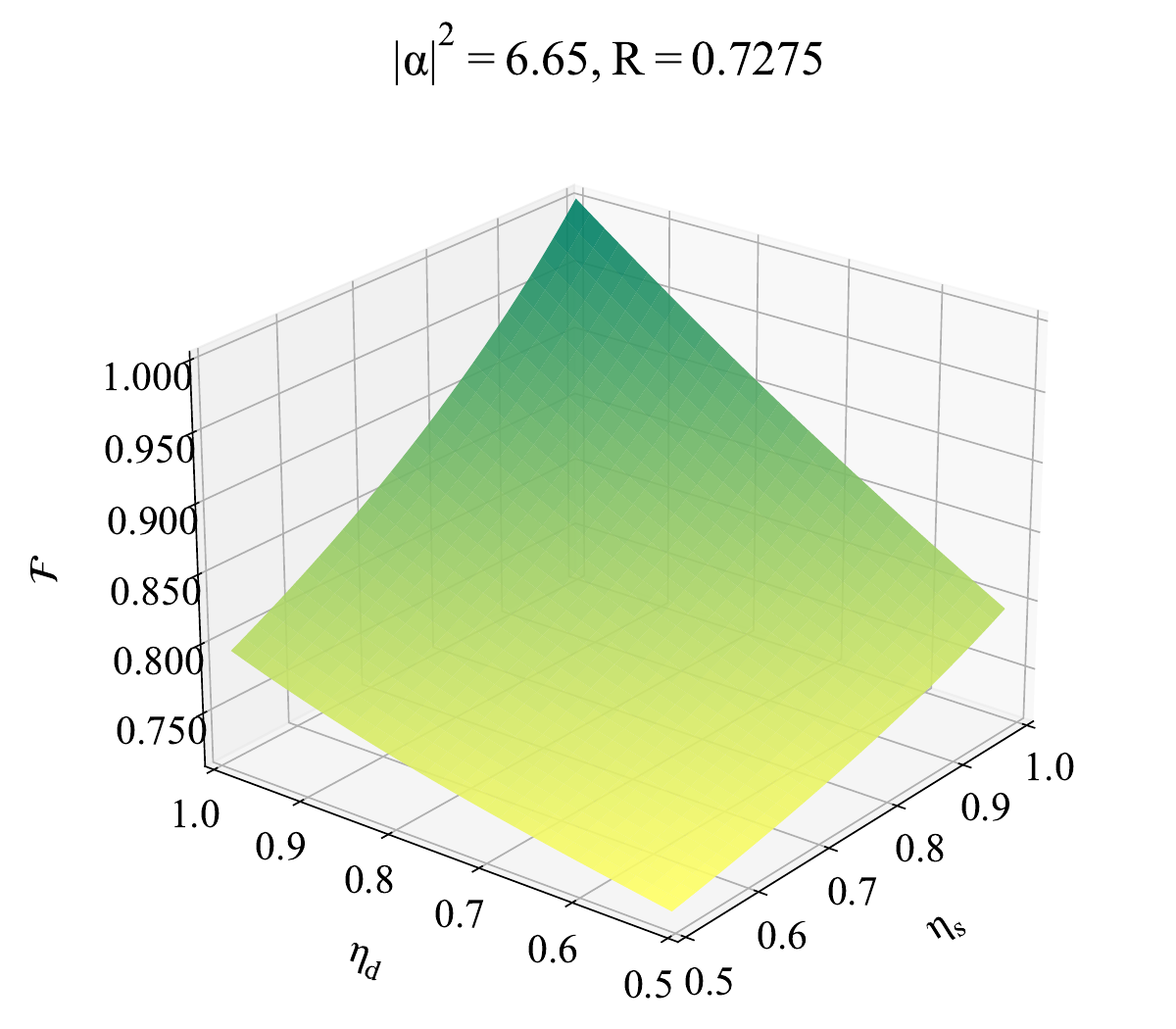}
         \subcaption[]{}
         \label{Fig.FidDQ51}
     \end{subfigure}
     \caption{(a), (b), (c) and (d) display the fidelity between ideal and realistic $\hbox{DQ}_{2}$, $\hbox{DQ}_{3}^{+1}$, $\hbox{DQ}_{4}^{+2}$ and $\hbox{DQ}_{5}^{+3}$ respectively.}
     \label{Fig.RCM}
\end{figure*}

To evaluate the experimental imperfections in the DQ preparation, it is necessary to measure the overlap between the ideal state and the experimentally realized state. This is achieved by calculating the fidelity $(\mathcal{F})$, which quantifies the similarity between the ideal density operator $(\hat{\rho}_I=\ket{\psi}_{nm}\bra{\psi})$ and the realized density operator $(\hat{\rho}_R)$. The fidelity is defined as 
\begin{equation*}   
    \mathcal{F} = \hbox{Tr}(\hat{\rho}_{I} \, \hat{\rho}_R).
\end{equation*}
The fidelity of unity means that the ideal and actual states match exactly, whereas the value below unity shows the deviation due to experimental imperfection. Fig. \ref{Fig.RCM}, illustrates the fidelity between the ideal and non-ideal $\hbox{DQ}$ for various $\abs{\alpha}^2$ and $R$. Fig. \ref{Fig.FidDQ21} sketches the realistic generation of optimal squeezer state $0.866\ket{0}+0.5\ket{1}$ with displacement from $\hbox{DQ}_{2}$. The higher fidelity achieved from the low efficiency of photon sources is observed in the creation of the displaced $0.9530\ket{0}-0.3030\ket{2}$ from $\hbox{DQ}_{3}^{+1}$, see Fig. \ref{Fig.FidDQ31}. This is because the photon sources considered mixed are similar to the resulting state.  

The non-ideal generation of the maximal squeezed states of $\hbox{DQ}_{4}^{+2}$ and $\hbox{DQ}_{5}^{+3}$, is presented in the figures \ref{Fig.FidDQ41} and \ref{Fig.FidDQ51}, respectively. The realistic generation of $\hbox{DQ}_{4}^{+2}$ is not significantly affected by the efficiency of photon sources. However, for the generation of the squeezed state $\hbox{DQ}_{5}^{+3}$, both $\eta_d$ and $\eta_s$ impact more on the non-ideal preparation. Based on the overall results, the fidelity of the generated states can be classified according to $\eta_s$. When the mixedness in the photon sources matches the characteristics of the desired superposition state, higher fidelity is observed. In all cases, fidelity is more strongly influenced by lower values of $\eta_d$ compared to $\eta_s$. An interesting observation is that the mixed nature of the photon source plays a significant role in facilitating the experimental realization of optimal squeezed states from DQ.
\subsection{Success Probability}
The success probability $(\hbox{S}_{p})$ estimates the probability of physical realization of a quantum state. For any quantum state, $\hbox{S}_{p}$ is the square of the normalization constant. The success probability of DQ is given by, 
\begin{equation}
    \hbox{S}_{p} = \Big[ \sum_{q=0}^n \abs{A_{nmq}}^2\Big]^{-1}.
\end{equation}
Table. \ref{tab:SP} presents the success probabilities $\hbox{S}_{p}$ for the respective maximal squeezed state that can be generated for the corresponding input number states of CM at $\eta_d=\eta_s=0.9$. The success probability of generating a superposed state from $\hbox{DQ}_{2}$ and $\hbox{DQ}_{3}^{+1}$, $\hbox{S}_{p}$ is close to $20 \%$. The success probability is around $15 \%$ for the optimal superposed state creation squeezed from $\hbox{DQ}_{4}^{+2}$ and $\hbox{DQ}_{5}^{+3}$.
\begin{table}[htbp]  
\centering
\caption{\bf The success probability and the Wigner negativity of optimal squeezed states generated at single photon detection.}
\begin{tabular}{ccccc}
\hline
$n$ & $\abs{\alpha}^2$ & $R$ &  $\hbox{S}_{p}$ $(\eta_d=\eta_s=0.9)$ & $W_{{N}}$ \\
\hline
$1$ & $3.05$ & $0.6000$ & $0.1898$ & $0.0298$ \\
$2$ & $5.45$ & $0.8175$ & $0.1699$ & $0.0580$ \\
$3$ & $6.00$ & $0.7650$ &  $0.1534$ & $0.0711$ \\
$4$ & $6.65$ & $0.7275$ &  $0.1392$ & $0.0784$ \\
\hline
\end{tabular}
  \label{tab:SP}

\end{table}
\section{Conclusions} \label{sec:CN}
Using linear optical components and photon detectors, conditional measurement schemes prompt the generation of various nonclassical states. In this study, we explored the output of CM as displaced qudits, with coherent states and photon number states as inputs. The results show that the input photon states limit the superposition up to its photon number, with the displacement. The detection alter the superposition coefficients. The superposition coefficients facilitate the desired finite superposition of number states by altering the coherent parameter and reflectivity of BS. The generation of novel quantum states like $A_{2m0}\ket{0}+A_{2m2}\ket{2}$ from $\hbox{DQ}_{3}^{\pm}$ which replicates the even coherent states with lower coherent amplitudes. 

The squeezing features of DQ are studied by employing quadrature squeezing and compared with the achievable squeezing in the finite superposition of Fock states. Fixing the detection event at one photon level, irrespective of the input number state, and carefully selecting the coherent state parameters and beam splitter reflectivity values lead to maximal squeezing, especially for lower number state inputs. Furthermore, the non-Gaussian properties of DQ are analyzed using HSD and $W_{N}$. From the observations, it is found that the non-Gaussianity of CM output is not enhanced than the input number states. We considered the effects of impure photon sources and detector inefficiencies to account for the experimental imperfections. The analysis revealed that the nonunit efficiency of the photon detector significantly degrades the squeezing properties of the output state. Moreover, the success probability of non-ideal generation of the optimal squeezed states from DQ is identified around $20\%$. 

\begin{acknowledgments}   
E. D and A. B. M Ahmed acknowledges the financial support of RUSA phase 2.0 of Madurai Kamaraj University.

\end{acknowledgments}

\bigskip

\bibliography{sample}

\end{document}